\documentclass[transaction]{IEEEtran}
\usepackage{caption}
\usepackage[nocompress]{cite}
\usepackage[cmex10]{amsmath}
\usepackage{graphicx}
\usepackage{picinpar}
\usepackage{authblk}
\usepackage{amsmath,amsfonts,amssymb}
\usepackage{array}
\usepackage{mdwmath}
\usepackage{mdwtab}
\usepackage{eqparbox}
\usepackage{stfloats}
\usepackage{subcaption}
\usepackage{ragged2e}
\usepackage{extarrows}
\usepackage{booktabs}
\usepackage{multirow}
\usepackage{etoolbox}
\usepackage{tabularx}
\usepackage{makecell}
\usepackage{longtable}
\usepackage[justification=centering]{caption}
\usepackage{siunitx}    
\usepackage{algorithmic}
\usepackage{xcolor}
\usepackage{bm}
\usepackage[margin=1.6cm]{geometry}
\usepackage{hyperref}
\usepackage{setspace}   
\usepackage{enumitem}   
\usepackage{longtable}
\usepackage{array}
\usepackage{stfloats}  
\usepackage{etoolbox}

\begin{document}

\newtheorem{lemma}{Lemma}
\newtheorem{corol}{Corollary}
\newtheorem{theorem}{Theorem}
\newtheorem{proposition}{Proposition}
\newtheorem{definition}{Definition}
\newcommand{\e}{\begin{equation}}
\newcommand{\ee}{\end{equation}}
\newcommand{\eqn}{\begin{eqnarray}}
\newcommand{\eeqn}{\end{eqnarray}}
\newcommand{\figsub}[2]{Fig.~\ref{#1}(\subref{#2})}

\renewcommand{\algorithmicrequire}{ \textbf{Input:}} 
\renewcommand{\algorithmicensure}{ \textbf{Output:}} 
\renewcommand{\raggedright}{\leftskip=0pt \rightskip=0pt plus 0cm}

\title{Toward Near-Space Communication Network in the 6G and Beyond Era}

\author{
\IEEEauthorblockN{Xinhua Liu\IEEEauthorrefmark{1}\IEEEauthorrefmark{2}\IEEEauthorrefmark{3}, Zhen Gao\IEEEauthorrefmark{1}\IEEEauthorrefmark{3}, Ziwei Wan\IEEEauthorrefmark{1}\IEEEauthorrefmark{2}, Zhonghuai Wu\IEEEauthorrefmark{1}\IEEEauthorrefmark{3}, Tuan Li\IEEEauthorrefmark{1}\IEEEauthorrefmark{3}, Tianqi Mao\IEEEauthorrefmark{1}\IEEEauthorrefmark{3}, Xiao Liang\IEEEauthorrefmark{1}\IEEEauthorrefmark{3}, Dezhi Zheng\IEEEauthorrefmark{1}\IEEEauthorrefmark{3}, Jun Zhang\IEEEauthorrefmark{1}\\
\IEEEauthorblockA{
\IEEEauthorrefmark{1}State Key Laboratory of Environment Characteristics and Effects for Near-space, Beijing 100081, China\\
\IEEEauthorrefmark{2}
School of Information and Electronics, Beijing Institute of Technology, Beijing 100081, China\\
\IEEEauthorrefmark{3}
School of Interdisciplinary Science, Beijing Institute of Technology, Beijing 100081, China}}
}

\maketitle

\begin{abstract}
Near-space communication network (NS-ComNet), as an indispensable component of sixth-generation (6G) and beyond mobile communication systems and the space-air-ground-sea integrated network (SAGSIN), demonstrates unique advantages in wide-area coverage, long-endurance high-altitude operation, and highly flexible deployment. This paper presents a comprehensive review of NS-ComNet for 6G and beyond era. Specifically, by contrasting satellite, low-altitude unmanned-aerial-vehicle (UAV), and terrestrial communications, we first elucidate the background and motivation for integrating NS-ComNet into 6G network architectures. Subsequently, we review the developmental status of near-space platforms, including high-altitude balloons, solar-powered UAVs, and stratospheric airships, and analyze critical challenges faced by NS-ComNet. To address these challenges, the research focuses on key enabling technologies such as topology design, resource and handover management, multi-objective joint optimization, 
etc., with particular emphasis on artificial intelligence techniques for NS-ComNet. Finally, envisioning future intelligent collaborative networks that integrate NS-ComNet with satellite-UAV-terrestrial systems, we explore promising directions. This paper aims to provide technical insights and research foundations for the systematic construction of NS-ComNet and its deep deployment in the 6G and beyond era.
\end{abstract}

\IEEEpeerreviewmaketitle

\section{Introduction}
\begin{figure}[h]
    \centering
  \includegraphics[width=0.5\textwidth]{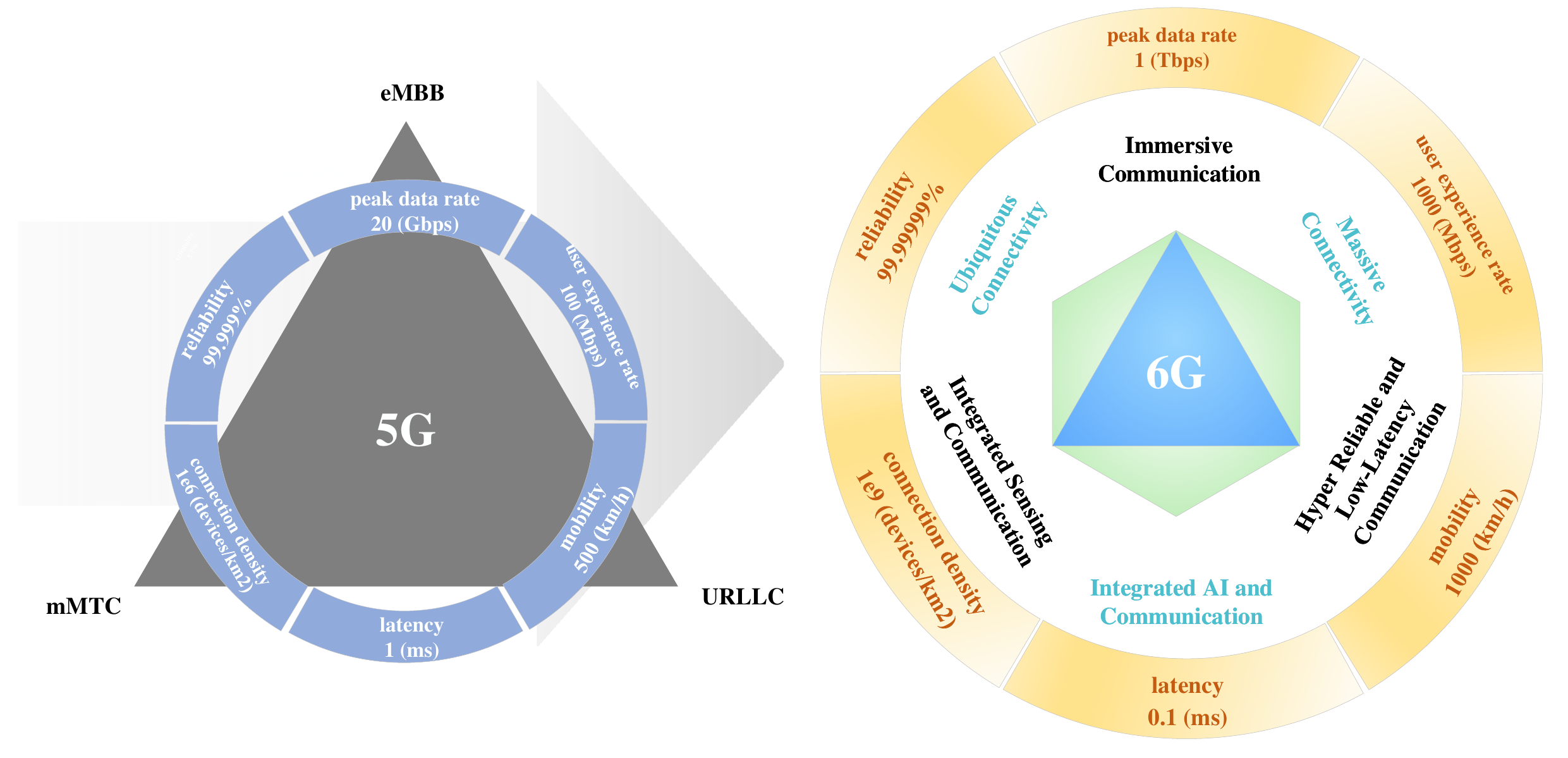}
    \caption{Performance improvements from 5G to 6G.}
    \label{5Gto6G}
\end{figure}

With the rapid growth in global communication demands and the accelerated development of the Internet of Things (IoT)~\cite{2023_Gao_LEO}, countries and organizations worldwide are actively exploring the sixth generation (6G) mobile communication technology. As shown in Fig.~\ref{5Gto6G}, evolving from the core capabilities of the fifth generation (5G) mobile communication technology, i.e., enhanced mobile broadband (eMBB), massive machine-type communications (mMTC), and ultra-reliable low-latency communications (URLLC)~\cite{ITU-RM2083}, 6G aims to enable immersive communication, ubiquitous connectivity, ultra-massive access, extremely reliable low-latency, and the convergence of communication with both sensing and artificial intelligence (AI)~\cite{ITU-2030}. 6G will support a fully integrated space-air-ground-sea architecture characterized by global coverage, multi-dimensional coordination, and intelligent resilience. Building upon the foundation of the Internet of Everything, 6G is committed to bridging the physical and digital worlds, overcoming the physical limitations of terrestrial networks, realizing truly ubiquitous global connectivity, and ultimately accelerating the transition toward an era of intelligent connectivity among all things~\cite{2023_Ke_IoE}.

\subsection{Background and Motivations}
\begin{figure*}[!ht]
    \centering
 \includegraphics[width=1\textwidth]{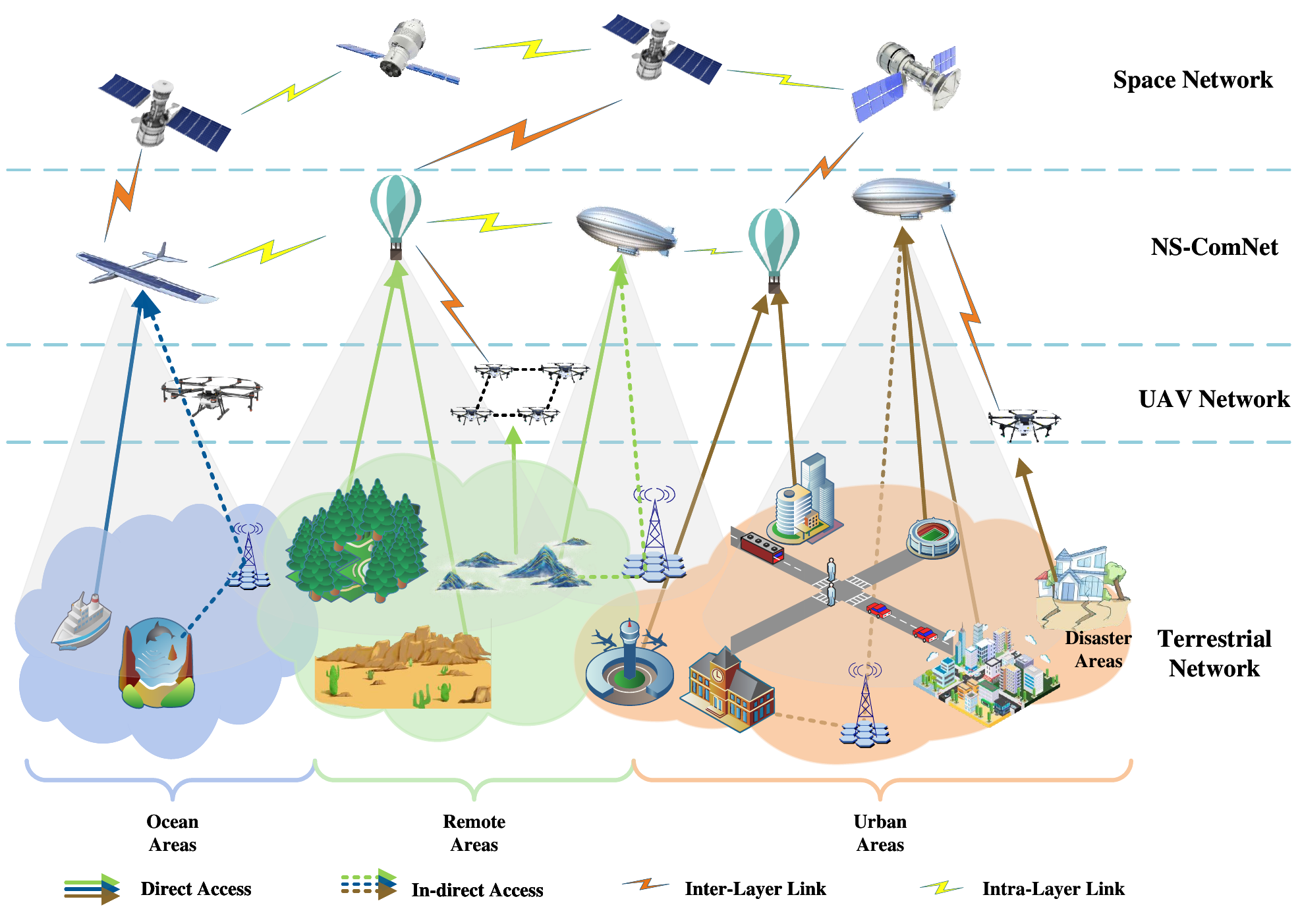}
    \caption{The space-air-ground-sea integrated network in the 6G and beyond era.}
    \label{SAGSIN}
\end{figure*}

Terrestrial networks deployed in developed and densely populated areas can achieve high economic efficiency and sustainable development at relatively low cost~\cite{2019_Huang_survey}. However, this inevitably leaves maritime, aerial, and sparsely populated terrestrial regions underserved in terms of high-speed and reliable communication. Even in dense urban environments, severe signal attenuation occurs due to the ``urban canyon effect" caused by building obstructions\cite{2015_urban_canyon}. In gatherings and emergency scenarios, temporary ultra-dense connectivity demands often exceed the elastic capacity of terrestrial networks. Moreover, natural disasters or deliberate attacks may paralyze local infrastructure, threatening communication security for critical services. The inherent physical constraints and architectural limitations of terrestrial systems are insufficient to realize the 6G vision of global coverage, ubiquitous connectivity, and diversified service requirements, highlighting the urgent need to extend the three-dimensional (3D) coverage of wireless networks. As illustrated in Fig.~\ref{SAGSIN}, the space-air-ground-sea integrated network (SAGSIN) fills this gap through (i) satellite-based wide-area coverage, (ii) near-space platforms (NSPs), also known as high-altitude platform stations (HAPS), enabling long-range line-of-sight (LoS) communication, and (iii) unmanned aerial vehicles (UAVs) for dynamic blind-spot compensation~\cite{2024_Gao_NSP}. In the foreseeable future, the space network, the near-space communication network (NS-ComNet), the UAV network, and the terrestrial network, either independently or collaboratively, will form a hierarchical and heterogeneous architecture. This deep integration of systems, technologies, and applications within SAGSIN represents a transformative paradigm for global connectivity and information exchange~\cite{2018_Liu_SAGSIN}. 

Satellite communications, as a critical component of SAGSIN, exhibit irreplaceable advantages in wide-area coverage and emergency response~\cite{2023_ying_LEO}. However, their inherent technical characteristics and deployment environments impose fundamental limitations that prevent satellites from independently fulfilling 6G requirements for global service provisioning~\cite{2021_Liu_satellite}. The latency and topological stability of space-based networks are constrained by orbital altitudes and satellite motion patterns, rendering them unsuitable for real-time-sensitive applications. Large-scale satellite constellations in low earth orbit (LEO) and frequent handovers exacerbate spectrum occupancy and orbital conflicts, jeopardizing the sustainable development of global communication systems~\cite{2023_zhou_LEO}. Satellite power budgets must be shared with propulsion and thermal control systems, drastically reducing available power for communication payloads. This restricts the deployment of large-scale antenna arrays, while power limitations, antenna size constraints, and spectral efficiency bottlenecks further hinder satellite networks' capacity to support high-density traffic demands. Additionally, the high costs of space-based infrastructure significantly impede their scalability for mass deployment.

Low-altitude UAVs, leveraging their flexible deployment and cost-efficiency advantages, have been widely adopted in logistics, agriculture, environmental monitoring, emergency rescue, security surveillance, scientific research, power line inspection, etc~\cite{2025_ran_UAV,2024_Gao_IoT}. However, their operational capabilities in complex terrains and wide-area missions are constrained by limitations such as a typical coverage radius under 50 km, an endurance generally below 4 hours, and a payload capacity below 20 kg~\cite{2019_Fotouhi_UAV}. Low-altitude operations also face technical bottlenecks, including positioning errors, limited wind resistance, and stringent airspace regulations. These factors degrade operational reliability, relegating UAVs to temporary coverage supplements rather than solutions for long-term continuous service requirements.

NS-ComNet deployed in 20-100 km above the Earth's surface effectively bridges the coverage and transmission distance gaps between space-based satellite networks and terrestrial networks. The near-space region above the stratosphere is characterized by stable atmospheric conditions, providing a physical environment for long-duration aerial operations~\cite{2022_Wu_CSI}. In terms of spatial coverage performance, a single NSP equipped with an extra-large-scale antenna array can establish a continuous service area with a diameter of hundreds of kilometers. Its top-down propagation paths significantly enhance signal penetration capabilities in complex terrains such as urban canyons. Regarding communication capacity, the synergistic application of terahertz (THz) frequency bands~\cite{2023_Ming_THz} and reconfigurable intelligent surfaces (RIS), combined with 3D beamforming and dynamic spectrum sharing technologies~\cite{2022_Tashiro_spectrum,2023_Ishikawa_spectrum}, enables substantial improvements in system spectral efficiency. This satiefies the 6G technical requirement for traffic density at the $\rm Tbps/km^2$ level. From an economic perspective, NSPs utilize stable stratospheric wind fields and high-efficiency solar power supply to achieve autonomous navigation and continuous operation with more than 99$\%$ of annual hover time. This results in significantly reduced operational costs compared to traditional satellite systems. As critical relay nodes in the SAGSIN, NSPs establish efficient heterogeneous integration channels between space-based satellite networks, low-altitude UAV networks, and terrestrial mobile networks. Leveraging their unique advantages, such as high transmission efficiency, extensive coverage, low communication latency, and strong anti-interference capability, NSPs effectively enhance end-to-end connectivity and enable seamless cross-domain communication.
    
\subsection{Contributions and Organization}
To elucidate the pivotal role of NSPs in SAGSIN, existing research has systematically reviewed their capabilities from multi-dimensional perspectives. In~\cite{2016_Flavi_survey}, d’Oliveira \textit{et al.} focused on aeronautical engineering, analyzing core technical bottlenecks in structural lightweight design, energy system optimization, low-altitude airspace management, and system reliability. X. Cao \textit{et al.} in~\cite{2018_Cao_survey} explored the advantages of NSPs in communication network architectures while establishing a research framework spanning application scenario adaptability, breakthrough paths for technical challenges, and channel modeling methodologies. J. Qiu \textit{et al.} in~\cite{2019_Qiu_survey} innovatively proposed a software-defined networking (SDN)-based heterogeneous collaborative architecture integrating high/low-altitude platforms with terrestrial networks, providing theoretical foundations for 3D integrated networking. S. C. Arum \textit{et al.} in~\cite{2020_Steve_survey} emphasized the technical superiority of NSPs in wide-area coverage enhancement and revealed the critical role of AI-empowered dynamic resource management in improving system efficiency. G. Karabulut Kurt \textit{et al.} in~\cite{2021_Karabulut_survey} further quantitatively demonstrated the underutilized performance potential of NSP systems through coordinated radio resource scheduling and physical-layer fusion technologies. Z. Xiao \textit{et al.} in~\cite{2022_Xiao_survey} systematically uncovered unique technical challenges in the near-space communications, particularly in terminal device design and computational task offloading, driven by emerging application demands.  T. Mao \textit{et al.} in~\cite{2022_Mao_survey} concentrated on THz band physical-layer technologies, constructing a comprehensive research framework encompassing application prospect analysis, key technical barrier resolution, and novel transmission model design. H. Liu \textit{et al.} in~\cite{2024_lhs_survey} provided a comprehensive survey of near-space communications, which serve as a critical enabler within SAGSIN. By examining deployment architectures, coverage capabilities, channel characteristics across various network layers, and air interface technologies, this work demonstrated the essential role of near-space communications in enhancing the connectivity, coverage, and transmission performance of conventional integrated network architectures.

Despite these reviews' comprehensive analyses of NSPs' current state and challenges covering from aeronautical engineering, networking architecture, resource management, and physical-layer perspectives, their research scope remains significantly limited. On the one hand, systematic studies on layered heterogeneous networks are scarce, with few works deconstructing the collaborative mechanisms and architectural paradigms between NSPs, satellites, UAVs, and terrestrial base stations from a holistic SAGSIN perspective. On the other hand, the transformative potential of intelligent technologies in reconfiguring NSP capabilities, particularly in AI-empowered transmission optimization, remains underexplored, lacking global-level analysis. This review aims to transcend the isolated optimization paradigms of traditional NSP research by systematically integrating cross-domain collaboration, (specifically referring to coordination among aerial, space, and terrestrial communication domains) along with intelligent endogenous mechanisms, thereby providing a holistic solution for NSP networking. The proposed framework not only bridges technical gaps in existing literature but also establishes theoretical foundations for transitioning NSPs from mere ``communication relays" to ``intelligent near-space based computing networks".

\begin{figure*}[h!]
    \centering   
    \includegraphics[width=1\textwidth]{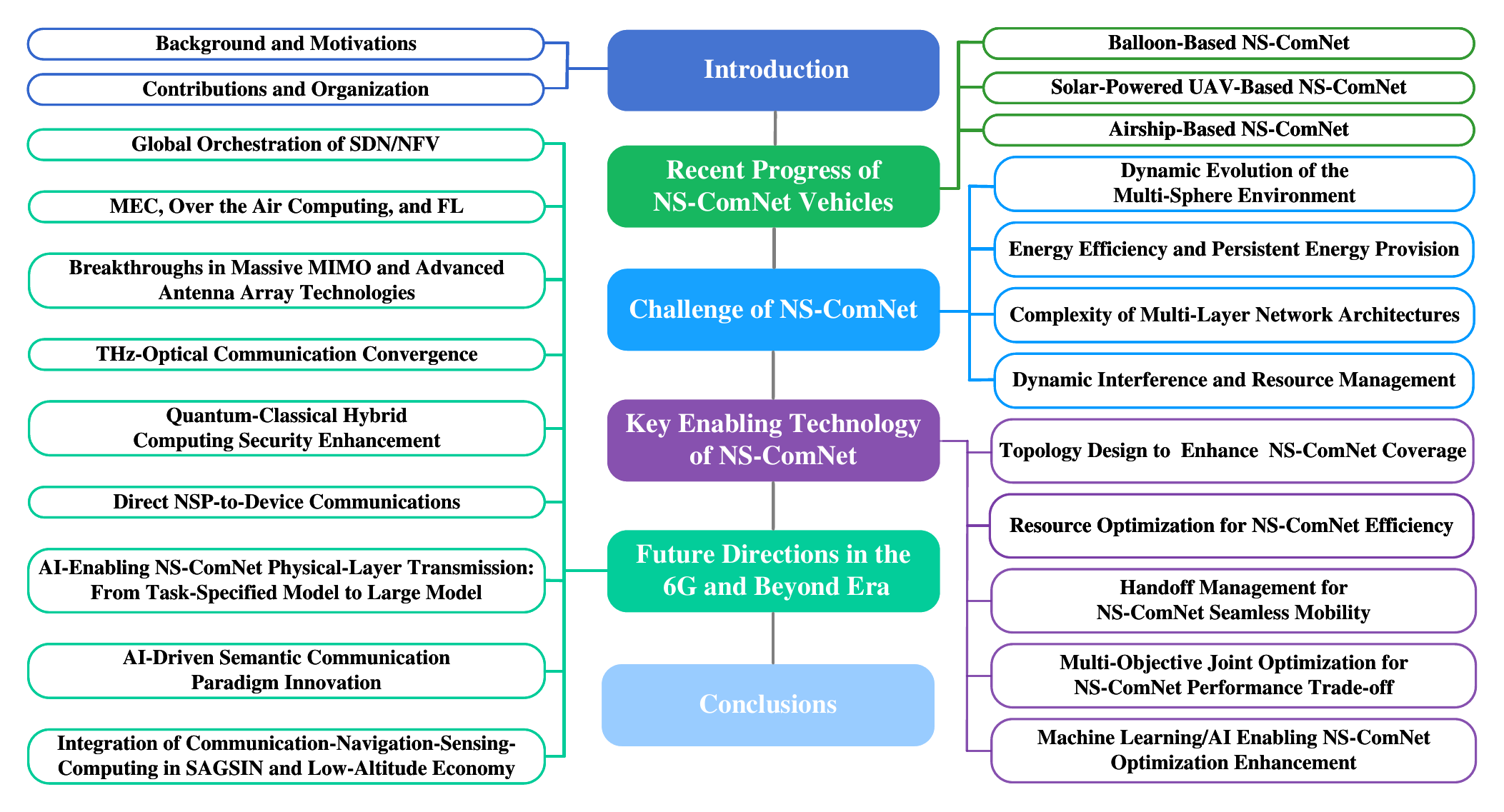}
    \caption{The structure of this paper.}
    \label{structure}
\end{figure*}

The structure of this paper is illustrated in Fig.~\ref{structure}. We begin by reviewing the current development status of NSPs, and then Section \ref{Challenge of NS-ComNet} analyzes the key challenges that constrain the advancement of near-space communications. In Section \ref{Key Enabling Technology of NS-ComNet}, we examine the collaborative mechanisms and architectural paradigms of SAGSIN from diverse perspectives, including topology design, resource allocation, handover management, joint optimization, and AI empowerment. Section \ref{Future Directions in the 6G and Beyond Era} presents a forward-looking discussion on cutting-edge technologies such as SDN/network functions virtualization (NFV), mobile edge computing (MEC), massive multiple-input multiple-output (mMIMO)~\cite{2024_Kui_CE} and novel antenna technologies, integrated THz and free-space optical (FSO) communications, hybrid quantum-classical algorithms, direct NSP-to-device communications, and large model-enabling physcial-layer transmission. The final section concludes the paper.

\section{Recent Progress of NS-ComNet Vehicles}
\label{Recent Progress of NS-ComNet Vehicles}
NS-ComNet vehicles, also known as NSPs, combine flexible deployment and cost-efficiency to rapidly respond to diverse mission requirements and establish emergency communication links. Their energy systems predominantly utilize solar power paired with high-efficiency energy storage, enabling continuous operation at near-space altitudes for months to years. This not only achieves seamless wide-area coverage, reduces base station handover frequency and operational costs but also fulfills stringent service continuity demands of modern communication systems. In terms of payload capacity, NSPs can carry communication equipment exceeding hundreds of kilograms due to their large structural size. This high-capacity characteristic supports multi-band composite communication services while enhancing network scalability through increased system redundancy. Considering coverage range, endurance duration, and operational stability, NSPs have emerged as a compelling solution for communication network deployment in remote regions and serve as core hubs for cross-domain coordination in the SAGSIN.

\begin{figure*}[!h]
\centering
\begin{minipage}[b]{0.48\linewidth}
	\subfloat[high-altitude balloon]{\label{balloon}
	\includegraphics[width=2.3in]{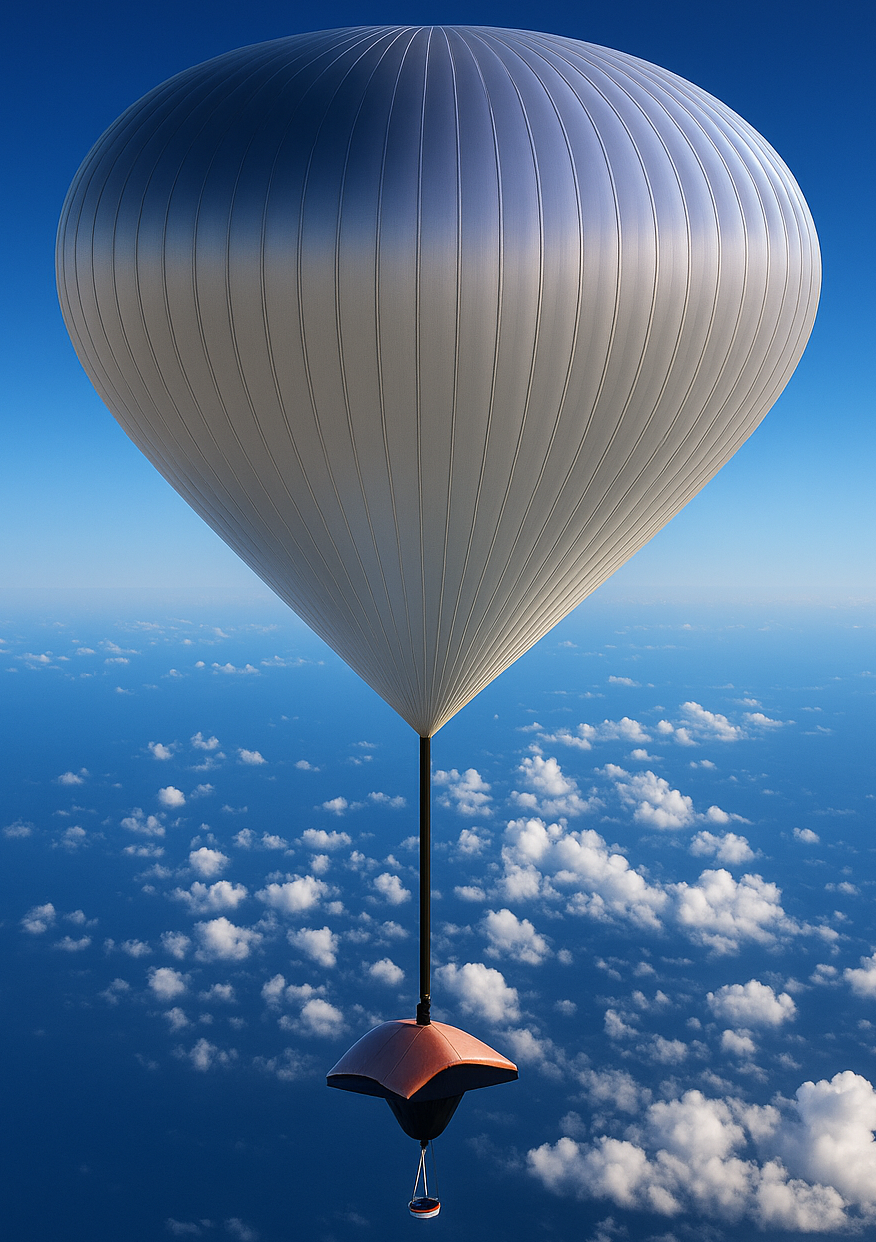}}
\end{minipage}
\hspace{-0.15\linewidth}
\begin{minipage}[b]{0.48\linewidth}
	\subfloat[solar-powered UAV]{\label{solar-powered}
	\includegraphics[width=3.1in]{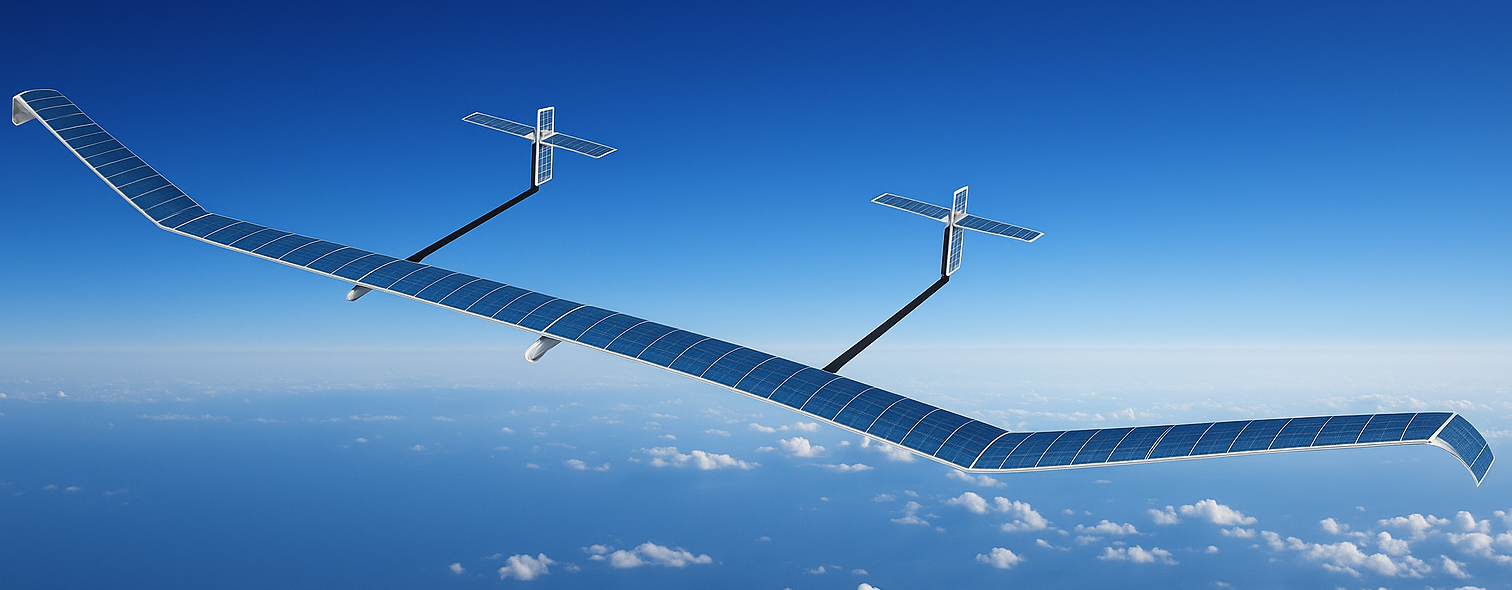}} \\  
	\subfloat[stratospheric airship]{\label{airship}
	\includegraphics[width=3.1in]{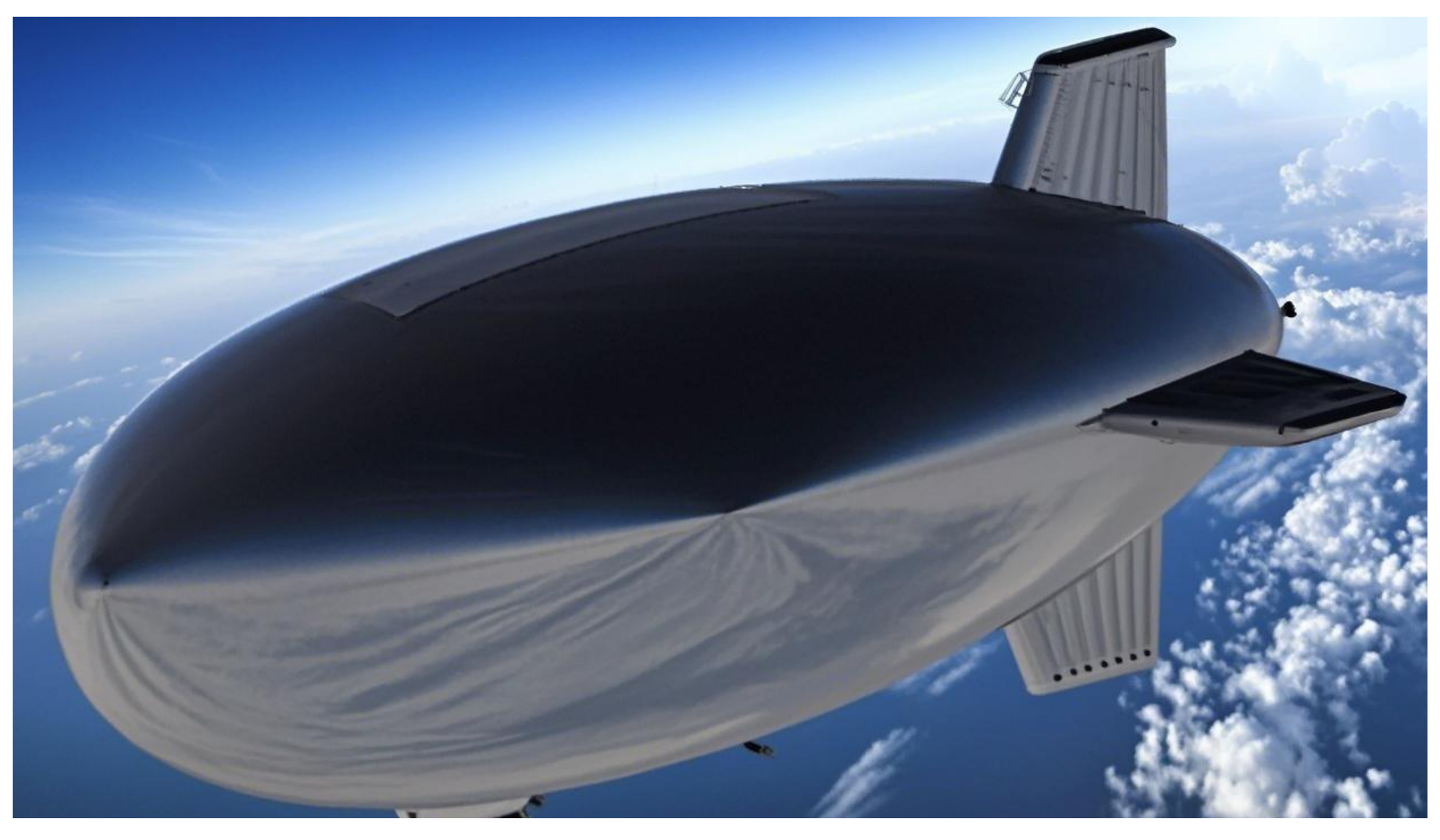}} 
\end{minipage} 
\caption{Representative NS-ComNet vehicles.}
\label{vehicles}
\end{figure*}

In addition, the near-space environment, characterized by weak turbulence intensity, atmospheric pressure at approximately 5$\%$ of sea-level conditions, and temperatures as low as \SI{-60}{\degreeCelsius}, enables near-space vehicles to maintain quasi-stationary hovering states for months to years even when carrying heavy payloads. As core carriers in NS-ComNet, typical NSPs primarily include high-altitude balloons, solar-powered UAVs, and stratospheric airships. High-altitude balloons, as shown in Fig.~\ref{vehicles}\subref{balloon}, achieve long-term aerial suspension via gas buoyancy principles, which enables cost-effective wide-area communication thanks to their passive hovering capability. However, their positioning accuracy remains constrained by atmospheric circulation effects. Solar-powered high-altitude UAVs, as shown in Fig.~\ref{vehicles}\subref{solar-powered}, employ aerodynamic flight modes, utilizing high-aspect-ratio wings and lightweight composite materials for sustained maneuverability. Hybrid energy systems combining solar cells and fuel cells support continuous operation for months, yet their payload capacities are typically limited to below 200 kg, with high-precision navigation systems increasing operational costs. Stratospheric airships, as shown in Fig.~\ref{vehicles}\subref{airship}, integrate active control mechanisms into buoyant architectures, featuring lengths exceeding 100 meters and flexible solar arrays atop their hulls. While electric propeller propulsion enables precise hovering, whereas complex aerodynamic designs face significant material durability challenges under intense stratospheric ultraviolet radiation.  Humanity’s exploration of near-space has progressively deepened, with advancing scientific understanding propelling near-space communication research into a new phase that equally prioritizes scientific exploration and application development, leveraging its advantages over satellite, terrestrial, and low-altitude networks.

\subsection{Balloon-Based NS-ComNet}
Led by Google X, the predecessor of Alphabet Inc., the Loon project was a representative aerostatic balloon-based high-altitude platform system implemented in the United States from 2011 to 2021~\cite{loon}. It aimed to achieve global communication coverage through stratospheric networking. The project established a mesh network architecture with the Loon SDN control system, ultimately serving over 300,000 users. Its final iteration demonstrated a stratospheric hovering duration of 312 days at altitudes of 18-23 km, with a single-node coverage radius of 40 km. Operational data revealed that the balloons accumulated over one million flight hours and a total range of 40 million km by 2019. These excellent performances set industry benchmarks for deployment scale and stability in high-altitude platforms and validating the engineering feasibility of long-term aerial operations. Although Project Loon was discontinued in 2021, its successful networking practices catalyzed industrial and academic advancements in NSP development. 

The high-altitude platform transport service introduced by Elevate (operated by Spain's Zero 2 Infinity since 2009), utilized the STRATOS vehicle for stratospheric payload delivery via buoyant balloon technology~\cite{zero2}. With a maximum payload capacity of 100-kg, it maintained 24-hour operational flights at altitudes of 18-22 km. This platform offered customizable combinations of altitude, duration, and payload mass, primarily serving technical validation and testing tasks for novel high-altitude communication equipment. Its standardized payload environment design enables repeatable experimental conditions for stratospheric technology development. Technical specifications and engineering insights from this platform have been incorporated as case studies in NSP system development.

\begin{table*}[h!]
\centering
\caption{Comparison of Representative NSP Projects}
\label{tab:HAP_compare}
%\resizebox{\textwidth}{!}{  % 自动缩放到整页宽度
\begin{tabular}{|%
  >{\centering\arraybackslash}m{2cm}|%
  >{\centering\arraybackslash}m{2.2cm}|%
  >{\centering\arraybackslash}m{4cm}|%
  >{\arraybackslash}m{7.5cm}|%
  }
\hline
\textbf{Project} & \textbf{Type} & \textbf{Company/ Country} & \multicolumn{1}{c|}{\textbf{Important Features}} \\
\hline

\textbf{Loon} & Balloon & Alphabet Inc. / USA &
\begin{itemize}[leftmargin=1em,itemsep=0.2em,topsep=0.2pt]
  \item It provided service to over 300,000 users.
  \item It can fly up to 312 days at an altitude around 18-23 km, with a 40 km coverage radius.
  \item It accomplished over one million flight hours, flying for a total of around 40 million km.
\end{itemize} \\
\hline

\textbf{Elevate} & Balloon & Zero 2 Infinity / Spain &
\begin{itemize}[leftmargin=1em,itemsep=0.2em,topsep=0.2pt]
  \item Its STRATOS vehicle can carry up to 100 kg for about 24 hours flight duration at altitudes between 18 km and 22 km.
\end{itemize} \\

\hline
\textbf{Jimu-1} & Balloon & AIR \textit{et al.}  / China &
\begin{itemize}[leftmargin=1em,itemsep=0.2em,topsep=0.2pt]
  \item It is  55-meter long and 19-meter high and has a volume of 9,060 cubic meters.
  \item It reached a maximum altitude of 9050 meters.
  \item It can maintain a stationary position for over 15 days, even at a temperature of \SI{-70}{\degreeCelsius}.
\end{itemize} \\
\hline

\textbf{X-station} & Airship & StratXX / Switzerland &
\begin{itemize}[leftmargin=1em,itemsep=0.2em,topsep=0.2pt]
  \item It achieved stable hovering at 21 km in the stratosphere, enabling a 1000-km-diameter coverage area.
  \item It supported continuous year-round operation with a 100 kg payload.
  \item It can establish a navigation network covering $10^6$ square km through three units.
\end{itemize} \\
\hline

\textbf{Yuanmeng} & Airship & Nanjiang South Space Sky Technology Co., Ltd. $\&$ Beihang University / China &
\begin{itemize}[leftmargin=1em,itemsep=0.2em,topsep=0.2pt]
  \item It achieved stable hovering at 20 km altitude for 48 days.
  \item It supported 100-300 kg payload and delivered 1000-watt payload support.
\end{itemize} \\
\hline

\textbf{Stratobus} & Airship & Thales Alenia Space / France &
\begin{itemize}[leftmargin=1em,itemsep=0.2em,topsep=0.2pt]
  \item It can support up to a 450-kg payload capacity and 5-year ultra-long endurance.
\end{itemize} \\
\hline

\textbf{SkyNet} & Airship & JAXA / Japan &
\begin{itemize}[leftmargin=1em,itemsep=0.2em,topsep=0.2pt]
  \item It consistes of several airships at an altitude at 20 km.
  \item Each airship would have about 200 meters length and could operate for up to 3 years covering a radius up to 100 km.
\end{itemize} \\
\hline

\textbf{BH-HAPs} & Airship & Linzhou Technology Co., Ltd. $\&$ Beihang University / China & 
\begin{itemize}[leftmargin=1em,itemsep=0.2em,topsep=0.2pt]
  \item It achieved stable hovering at 20km altitude for months. 
  \item It can take 5G/6G air-ground integrated communication system and can be used for maritime and sparsely populated arreas.
\end{itemize} \\
\hline

\textbf{SunGlider} & Solar-powered UAV & HAPSMobile Inc. / Japan \& USA &
\begin{itemize}[leftmargin=1em,itemsep=0.2em,topsep=0.2pt]
  \item Through five stratospheric test flights, it had validated its high-altitude communication relay capabilities.
\end{itemize} \\
\hline

\textbf{Zephyr S} & Solar-powered UAV & Airbus Defense and Space / UK &
\begin{itemize}[leftmargin=1em,itemsep=0.2em,topsep=0.2pt]
  \item It has a 25-meter wingspan and its operational altitude exceeds 18 km.
  \item It has a hybrid power system combining thin-film solar cells with lithium-ion batteries, delivering continuous 250-watt-class payload support.
  \item It can broadcast at a rate of 100 Mbps with a payload of up to 12 kg and can fly continuously for 100 days.
\end{itemize} \\
\hline

\end{tabular}
%}
\end{table*}

Between January 2019 and May 2022, the ``Jimu-1" tethered balloon, also known as an aerostat, was independently developed by the Aerospace Information Research Institute of the Chinese Academy of Sciences (AIR) and other institutions, and successfully conducted ten atmospheric scientific observation missions~\cite{Jimu}. Reaching a maximum altitude of 9,050 meters-surpassing Mount Everest, it set a world record for the highest in-situ atmospheric observation by an aerostat. Constructed from composite fabric materials, ``Jimu-1" is 55 meters in length, 19 meters in height, and has a volume of 9,060 cubic meters. It was lifted by buoyant gas and controlled via ground-based tethering systems, capable of maintaining a stationary position in unrestricted airspace for over 15 days, even in temperatures as low as \SI{-70}{\degreeCelsius}.
\subsection{Solar-Powered UAV-Based NS-ComNet}
Since its establishment in 2017, HAPSMobile has focused on developing the SunGlider platform, which is a solar-powered UAV with a 78-meter wingspan~\cite{HAPSMobile}. Through five stratospheric test flights, the platform validated its high-altitude communication relay capabilities. Leveraging communication payloads enhanced by Project Loon's technological improvements, HAPSMobile collaborated with the Loon team to develop stratospheric long term evolution (LTE) services, targeting commercial deployment by 2023. 

The aerodynamics-based NSP system named Zephyr S, developed by Airbus Defence and Space~\cite{Zephyr}, aimed to provide connectivity for populations in globally remote regions through stratospheric communication relays. Featuring a 25-meter wingspan and operational altitude exceeding 18 km, this platform employed a hybrid power system combining thin-film solar cells with lithium-ion batteries, delivering continuous 250-watt class payload support. Its technical indicators include the data broadcasting capability at a rate of 100 Mbps, the maximum payload capacity of 12 kg, and the endurance performance of 100 consecutive flight days in a single mission~\cite{2021_Karabulut_survey}. Flight data since 2013 have demonstrated that Zephyr S has established the continuous station-keeping flight record for aerodynamic NSPs, completing 25 days of sustained stratospheric navigation during initial operations, which is a critical validation of persistent aerial platform feasibility. 

\subsection{Airship-Based NS-ComNet}
The X-station high-altitude airship platform, developed by Swiss company StratXX since 2005~\cite{stratXX}, employed aerodynamic lift principles to achieve stable hovering at 21 km in the stratosphere, enabling a 1,000-km-diameter coverage area. This platform integrated multi-modal communication systems compatible with television signal broadcasting, radio transmission, cellular mobile communications, VoIP voice services, and remote sensing data transmission. Powered by a solar-battery hybrid energy system, it supported continuous year-round operation with a 100 kg payload. Through coordinated networking of three X-station units, a localized enhanced navigation network covering $10^6$ square km can be established, delivering sub-meter-level global position system positioning accuracy for designated regions. This technological approach validated the engineering feasibility of high-altitude airships in SAGSIN and navigation architectures.

In 2015, the ``Yuanmeng" airship, jointly developed by Nanjiang South Space Sky Technology Co., Ltd. and the team led by Prof. Z. Wu at Beihang University, completed the world’s first near-space airship flight featuring sustained propulsion, controllable navigation, and reusability~\cite{ym}. Powered by solar energy and lifted by helium, it reached an altitude of approximately 20 kilometers, carrying payloads for broadband communication, data relay, high-definition observation, and aerial situational awareness. The 48-hour flight demonstrated the feasibility of long-endurance and multi-mission operations in near-space environments.

In Europe, the Stratobus airship project led by Thales Group ~\cite{Stratobus}, funded by the European Space Agency, advances the development of hydrogen fuel cell and thin-film solar hybrid power systems. Targeting a 450-kg payload capacity and five-year ultra-long-endurance, its 220-meter airship design and ion-thruster positioning system have entered the engineering verification phase.

The SkyNet project was an early stratospheric platform communication initiative led by the National Aerospace Laboratory of Japan (NAL), which later became part of the Japan Aerospace Exploration Agency (JAXA)~\cite{SkyNet}. Conducted from 1998 to 2005, the project aimed to support future high-speed communication links by deploying airships at an altitude of 20 km in the stratosphere. Each airship was envisioned to be approximately 200 meters in length, capable of operating for up to three years, and providing a ground coverage radius of about 100 km. Although several phases of the project were successfully completed, the program was ultimately terminated due to funding limitations.

In recent years, the ongoing advancement of communication technologies has significantly contributed to the development of NS-ComNet.
Additionally, improvements in solar cell efficiency, innovations in lightweight composite materials, and breakthroughs in autonomous avionics systems and antenna technologies have further accelerated its emergence as a critical component for aerial networking. With the shortening of technology iteration cycles and the large-scale adoption of cost-effective materials, the economic feasibility of NSPs is poised to be significantly enhanced in future network deployments.

Table ~\ref{tab:HAP_compare} lists some popular past and recent projects along with their classification and key features. 

\section{Challenges of NS-ComNet}
\label{Challenge of NS-ComNet}
Although numerous national research initiatives and commercial projects have conducted engineering validations across various platform types, NSPs still face multiple constraints across various aspects in promoting cross-domain collaborative networking within SAGSIN. These challenges primarily arise from the dynamic and complex nature of both internal and external environments, as well as inherent system-level technical limitations.

\subsection{Dynamic Evolution of the Multi-Sphere Environment}
Research on multi-sphere coupling faces fundamental challenges. The cross-scale energy interactions between exogenous and endogenous sources are highly nonlinear, and the resultant multi-physical-field coupling is exceptionally complex~\cite{li2024direct}. First, the energy and mass transfer across atmospheric, ionospheric, and magnetospheric spheres exhibit cross-spatiotemporal scale properties that cannot be fully characterized by single-sphere models or traditional linear parameterization methods~\cite{ganse2025vlasiator}. Second, real-time modeling of correlations between exogenous perturbations (such as solar energetic particle streams and cosmic rays) and endogenous responses (such as atmospheric density gradients and electron concentration fluctuations) is hindered by inconsistencies in spatiotemporal data benchmarks and incompatibility of physical dimensions. Furthermore, uncertainties in cross-sphere energy transfer pathways impose stricter requirements for real-time environmental sensing and early warning systems~\cite{kang2024multi}.

\begin{figure}[h!]
    \centering   
    \includegraphics[width=0.5\textwidth]{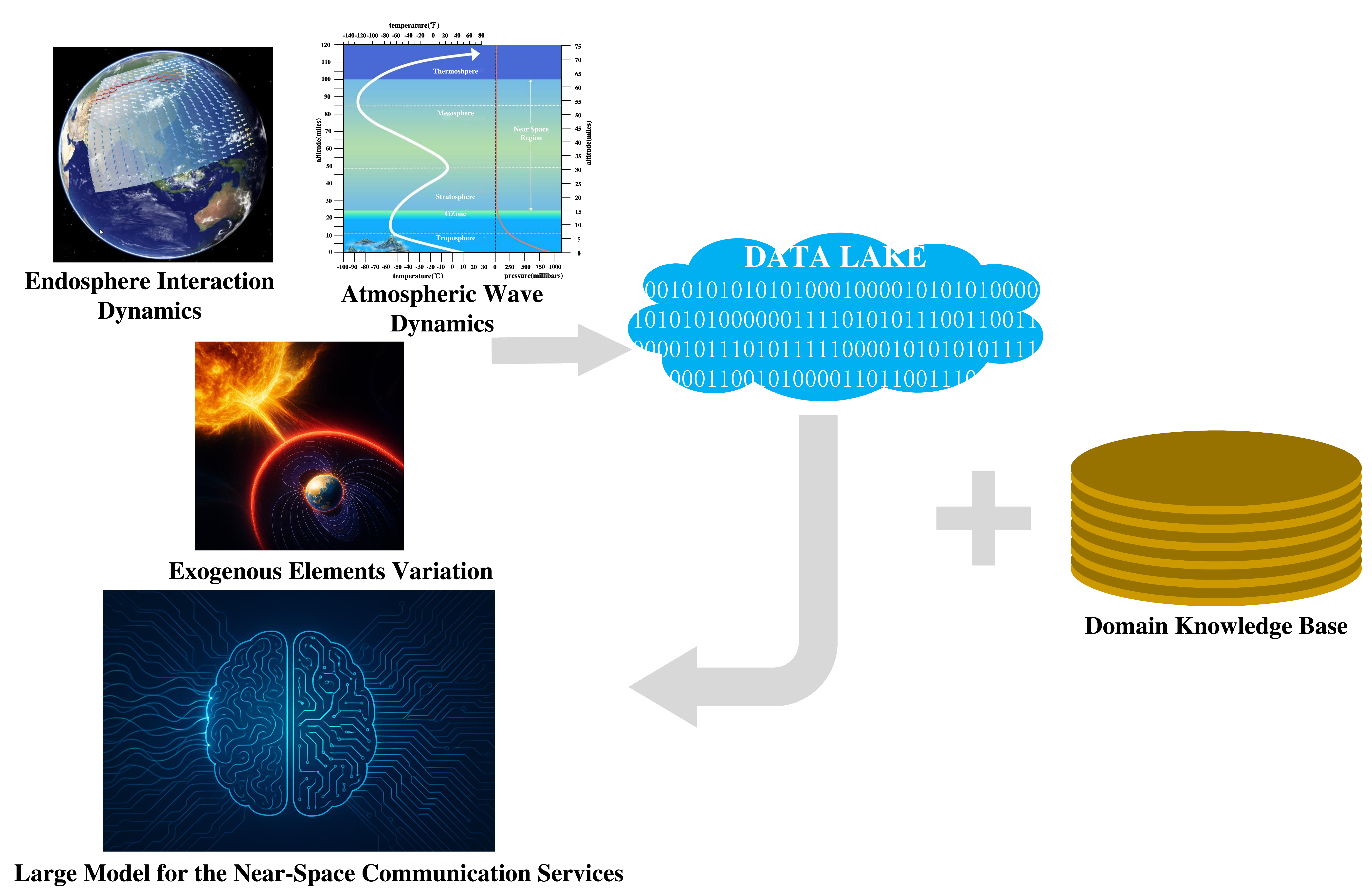}
    \caption{Core challenges of dynamic evolution of the multi-sphere environment.}
    \label{challenge}
\end{figure}

Meanwhile, constructing a large model for the near space, which encompasses thermosphere, mesosphere, stratosphere, and lower atmospheric layers, relies on integrating heterogeneous, multi-source datasets~\cite{2025_LM_Jing}. These datasets are continuously aggregated and preprocessed in a data lake infrastructure that supports large-scale spatiotemporal fusion and model training~\cite{zhang2024datafusion}. On this basis, a domain knowledge base encompassing atmospheric wave dynamics, endosphere interaction physics, and exogenous element variations can be constructed to enhance interpretability and generalization performance of AI-assisted inversion engines. Furthermore, 3D visualization of cross-sphere energy redistribution requires processing massive heterogeneous data streams, yet existing edge computing architectures fail to meet disaster response demands in terms of energy efficiency and computational latency.   The cross-temporal and cross-spatial dynamic feedback effects arising from multi-sphere coupling fundamentally constrain the modeling accuracy of multi-physical-field interactions in the near-space environment~\cite{zheng2024path}. This leads to systemic bottlenecks impacting dynamic vehicle response prediction, electromagnetic plasma control, and blackout communication recovery. As illustrated in Fig.~\ref{challenge}, these challenges highlight the core contradiction in near-space environmental research: how to construct a full-chain coupling evolution model that balances physical interpretability and engineering applicability under the triple constraints of exogenous perturbation randomness, endogenous response nonlinearity, and cross-sphere correlation complexity.

\subsection{Energy Efficiency and Persistent Energy Provision}
NSPs' energy management faces multi-dimensional technical challenges, with the core contradiction rooted in the significant gap between long-term stratospheric station-keeping requirements and the efficiency, capacity, and environmental adaptability of existing energy systems. At the technical architecture level, the divergent energy demands caused by payload functional variations across different NSP types remain particularly prominent. Current mainstream solutions relying on solar-battery systems show critical limitations: (i) gallium arsenide batteries exhibit only 18$\%$-22$\%$ daily energy conversion efficiency in the stratosphere, and it will be significantly reduced by seasonal solar incidence angle variations and high-intensity radiation; (ii) lithium-sulfur batteries suffer over 40$\%$ effective capacity degradation under extreme low-temperature conditions, failing to meet year-round uninterrupted operation demands~\cite{liu2023challenges,2020_Arum_Energy}.

The three aspects of energy supply and energy storage optimization proposed by academia still faces technical constraints: (i) In terms of energy, hybrid power systems combining hydrogen fuel cells and perovskite-silicon tandem cells are constrained by low-temperature start-up issues inherent in hydrogen energy and radiation stability problems of perovskite materials~\cite{Perovskite/Silicon}; (ii) In terms of load, reflective solar energy systems leverage surface segmentation strategies to maximize the utilization of solar irradiance and enhance energy distribution flexibility. However, their high reflectivity and the discontinuities introduced by structural segmentation may interfere with antenna radiation patterns, thereby degrading the directional performance of communication beams; (iii) In terms of control, although SDN and NFV are introduced to enable global resource orchestration and achieve decoupling between hardware and software, their implementation often incurs substantial hardware virtualization overhead. While such decoupling enhances system flexibility and improves resource utilization efficiency, the associated overhead ultimately offsets the expected gains in energy efficiency. Furthermore, scenarios such as stratospheric turbulence, ionospheric disturbances, and peak power consumption in AI-driven tasks urgently require the establishment of energy optimization models spanning multiple temporal scales.

\subsection{Complexity of Multi-Layer Network Architectures}
The integration of space-air-ground-sea vertical heterogeneous networks (VHetNets) faces systemic challenges arising from multi-dimensional heterogeneities, with complexity primarily stemming from the intertwined effects of four key heterogeneous dimensions~\cite{2023_Xu_SAGISN}. In coverage scalability, multi-orbit satellites, NSPs, low-altitude UAVs, and terrestrial base stations demonstrate hierarchical characteristics of global coverage, wide-area coverage, local coverage, and fixed-point coverage, respectively. In mobility patterns, nodal entities across layers exhibit different movement modes including geostationary orbits, quasi-stationary hovering, dynamic cruising, and fixed deployment. In energy provisioning, continuous solar power supply, finite fuel cell endurance, and stable terrestrial power supply form multi-level energy-constrained systems. In service capability, network layers present orders-of-magnitude disparities in transmission rates, latency characteristics, and connection densities. The multiple aspects directly induces nonlinear dynamic evolution of network topology.
\begin{figure}[h!]
    \centering   
    \includegraphics[width=0.5\textwidth]{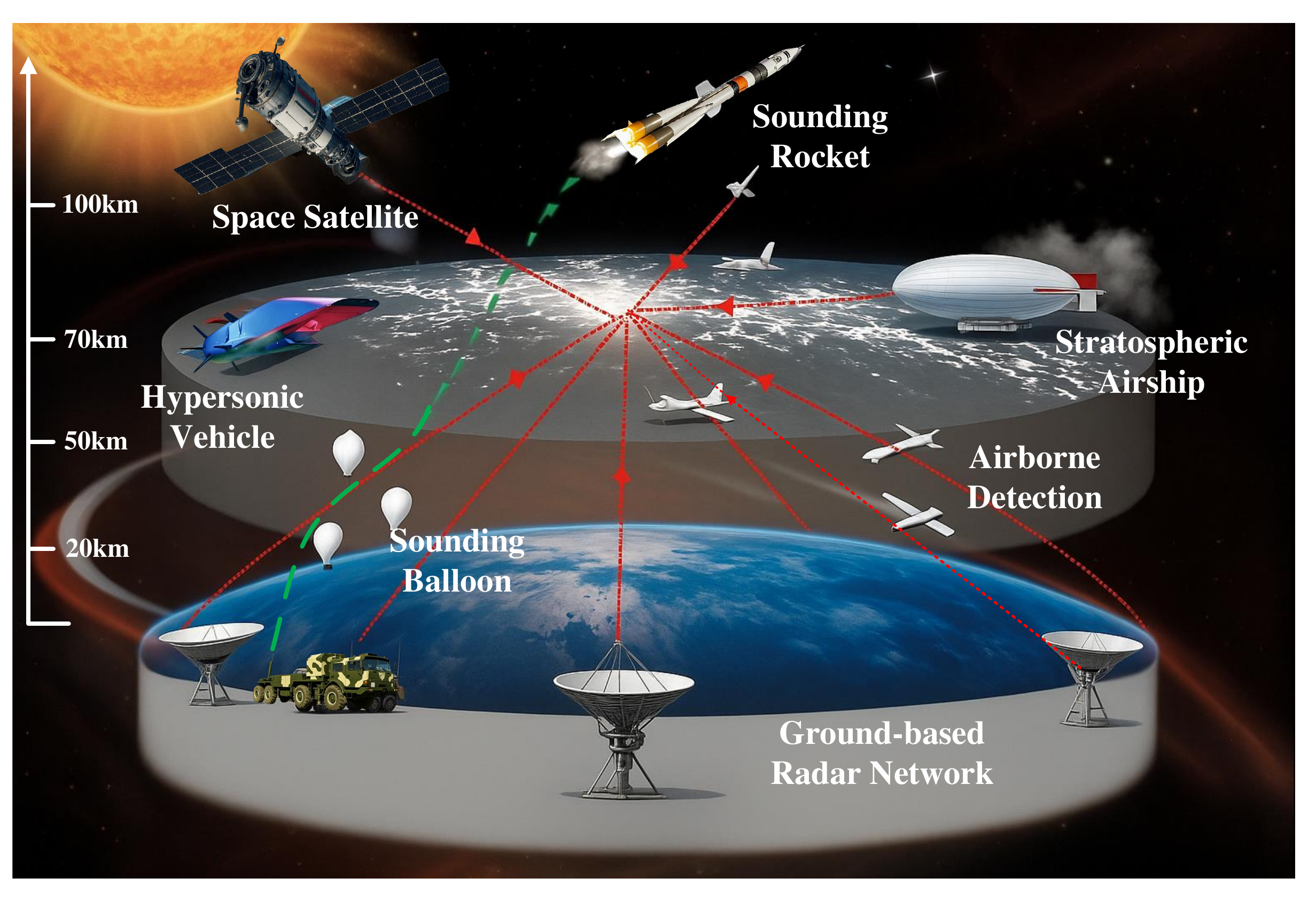}
    \caption{Multi-platform integrated sensing and communication architecture in the near-space environment.}
    \label{VHetNets}
\end{figure}

A representative example of such multi-layer complexity is illustrated in Fig. \ref{VHetNets}, which depicts a comprehensive near-space sensing and communication system integrating diverse platforms such as space satellite, stratospheric airship, hypersonic vehicle, sounding rocket, sounding balloon, and ground-based radar network. These heterogeneous nodes collectively operate across a wide vertical range, covering altitudes ranging from 20 to over 100 km, requiring precise cross-layer coordination for situational awareness, environmental prediction, and mission assurance. The layered deployment of sensing and communication assets reflects the challenges of harmonizing variable coverage ranges, temporal dynamics, and communication latencies. Moreover, the integration of data streams from multiple altitudes and modalities imposes strict demands on the network’s real-time processing and adaptive control capabilities. This exemplifies the inherent architectural complexity in achieving seamless interoperability and reliable service delivery in VHetNets.

Although SDN and NFV technologies provide theoretical frameworks for dynamic control, the heterogeneity of the protocol stack still leads to a delay bottleneck in cross-layer signaling interaction on the order of hundreds of milliseconds and the risk of interface adaptation failure. Current research focuses on constructing a trinity architecture of ``dynamic perception-intelligent decision-resilient reconfiguration", with key breakthroughs including fault-tolerant networking mechanisms driven by high-dimensional network state awareness, deep reinforcement learning (RL)-based robust transmission protocols for weak connectivity~\cite{2023_Gao_RL}, and cross physical-network-application layer heterogeneous data co-transmission frameworks~\cite{zhang2018deep}.
\subsection{Dynamic Interference and Resource Management}
The co-channel and heterogeneous interferences in multi-layer networks manifest an intertwined superposition within 3D airspace, with spatiotemporal dynamics far exceeding those of traditional terrestrial networks, driving fundamental paradigm shifts in interference control methodologies. The dynamic mMIMO beams of stratospheric NSPs combined with LEO satellite wide-area coverage frequently induce cross-layer beam overlapping. When these overlaps are superimposed with the cumulative co-channel interference effects from densely deployed UAV swarms, complex interference superposition fields are formed. Conventional interference coordination techniques such as game theory-based power control show significant limitations in multi-dimensional dynamic environments~\cite{wang2014game}, necessitating the construction of multi-objective joint optimization frameworks integrating spectrum-energy-computing resources. Although AI technologies like deep learning have demonstrated computational efficiency advantages~\cite{2017_Sun_AI}, there are three critical constraining factors: (i) incomplete differentiated quality of service (QoS) modeling for heterogeneous services such as URLLC and mMTC, (ii) underdeveloped dynamic coordination mechanisms between solar-powered energy supply and communication loads, and (iii) the lagging dual-connectivity handover protocol and mobility management framework in scenarios with high millimeter-wave (mmWave) attenuation. Moreover, novel service scenarios require breaking through traditional evaluation systems. For example, the quantitative modeling of the age of information and the quality of experience has not yet been integrated into the resource allocation mechanism of NSPs, and the channel models and access strategies for vertical industries such as cargo drones still need to be improved.

As strategic nodes within SAGSIN architecture, NSPs play irreplaceable roles in emergency-response communications, wide-area IoT coverage, and the low-altitude economy. These benefits derive from their large coverage radius, multi-seasonal station-keeping capability, and ability to dynamically reconfigure network topology. Nevertheless, platform stability under extreme temperature-pressure conditions, the robustness of cross-medium communication links, and the lack of unified space-air-ground standards continue to impede the systematic advancement of the SAGSIN architecture.
\section{Key Enabling Technologies of NS-ComNet}
\label{Key Enabling Technology of NS-ComNet}
\begin{figure}[h]
    \centering
    \includegraphics[width=0.5\textwidth]{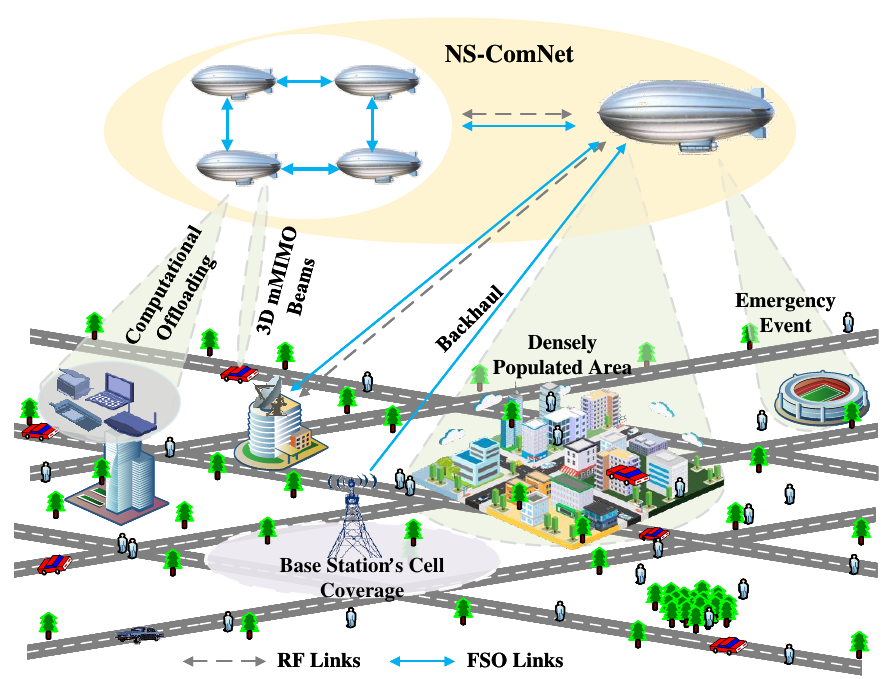}
    \caption{Typical use cases of NS-ComNet.}
    \label{NS-ComNet}
\end{figure}

A hierarchical cross-domain network architecture constructs a multi-dimensional infrastructure by integrating satellites, NSPs, UAVs, and terrestrial networks, enabling space-air-ground coverage and on-demand service. Inter-node connectivity is provided through radio frequency (RF) links or FSO links. The essence of VHetNets lies in functional decoupling through differentiated deployments: (i) the satellite layer provides ubiquitous global connectivity and inter-satellite transmission, (ii) NSPs provide wide-area coverage and backbone relaying, (iii) UAVs enhance hotspot capacity via agile mobility, and (iv) terrestrial base stations ensure high-density user access. Optimization of the NS-ComNet faces challenges including cross-domain node heterogeneity, dynamic topologies, and high propagation latency, necessitating breakthroughs beyond conventional single-layer optimization frameworks. This requires integrating coupled variables such as spectrum allocation, interference suppression, and mobility management into unified mathematical models to establish a joint optimization system spanning multi-dimensional resources. Such systems dynamically balance spectral efficiency, energy efficiency, and deployment costs while supporting critical missions, e.g., outdoor small-cell backhaul, emergency response, computational offloading, wide-area connectivity, and intelligent transportation system coverage. These missions must be sustained under complex electromagnetic environments and dynamic topologies, as depicted in Fig.~\ref{NS-ComNet}.

This section systematically analyzes design principles and technical features of representative NS-ComNet topologies, with a focus on key mechanisms such as dynamic spectrum sharing algorithms. These solutions aim to balance heterogeneous resource competition and guarantee service continuity, thereby providing architectural-level theoretical support for subsequent research endeavors.
\subsection{Topology Design to Enhance NS-ComNet Coverage}
\begin{figure}[h]
  \centering
  \begin{subfigure}[b]{1\columnwidth}
    \includegraphics[width=\columnwidth]{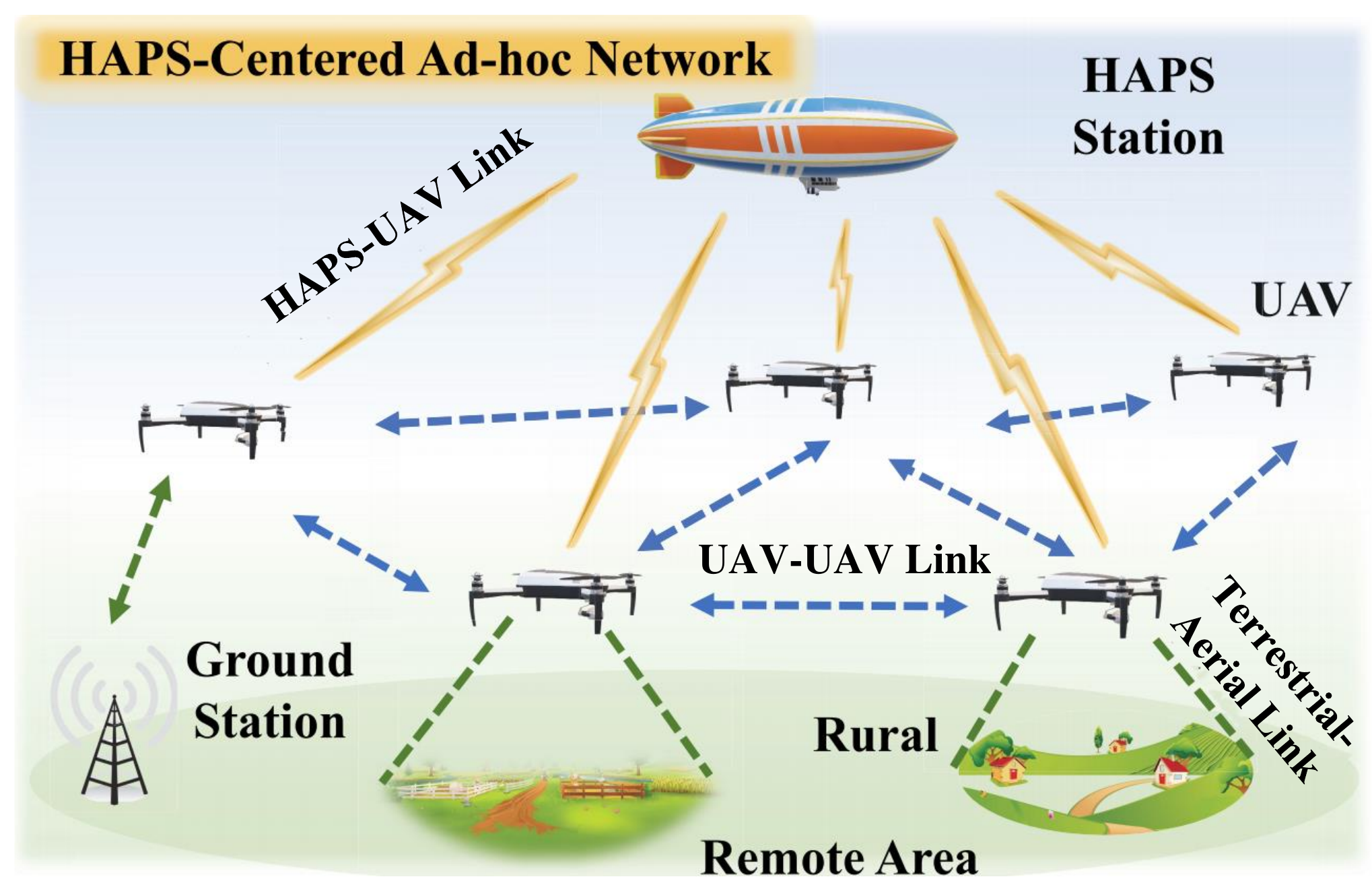}
    \caption{HAPS-centered ad-hoc network}
    \label{NSP_center}
  \end{subfigure}
  \vspace{0.1cm} 
  \begin{subfigure}[b]{1\columnwidth}
    \includegraphics[width=\columnwidth]{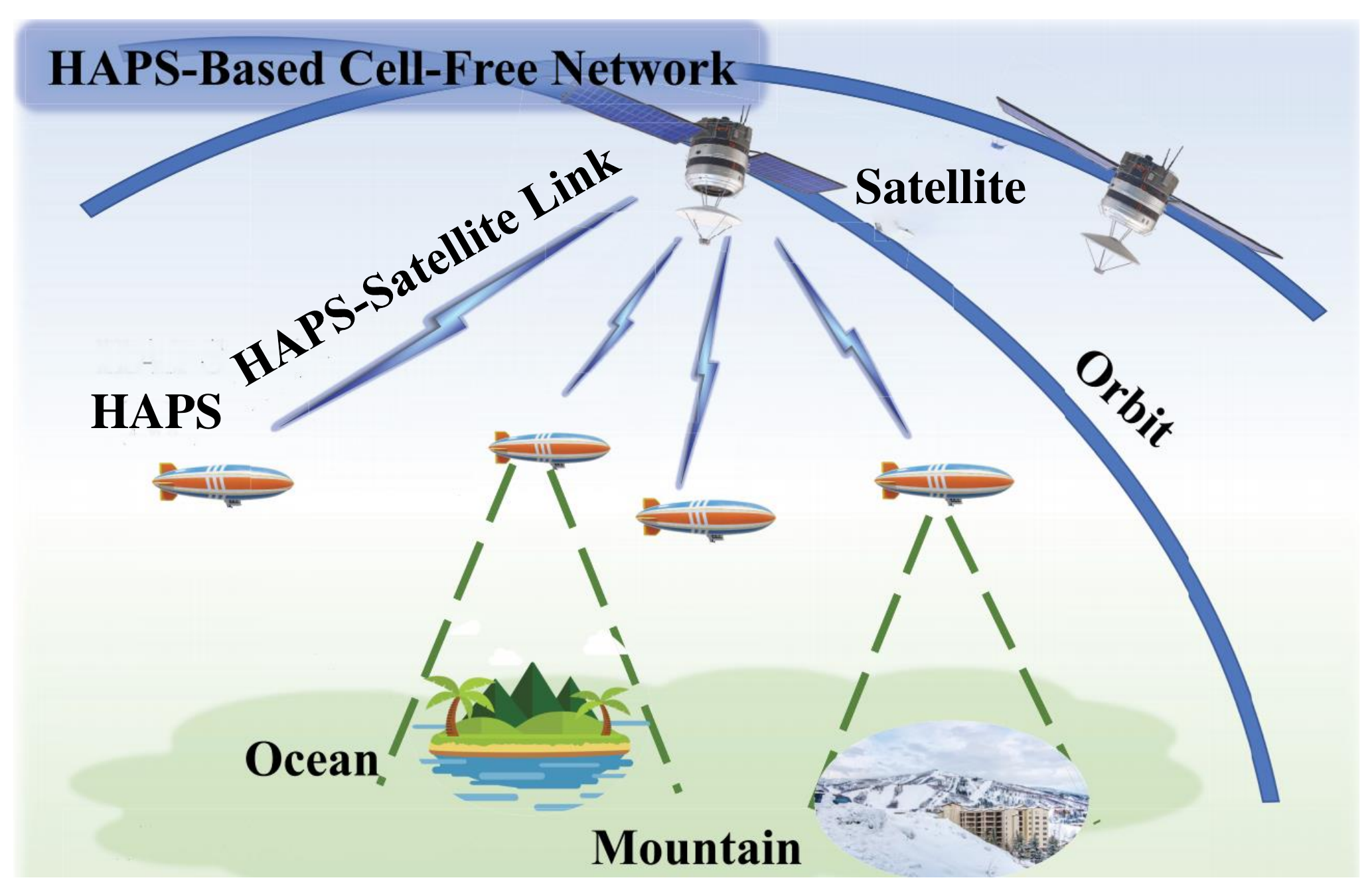}
    \caption{HAPS-based cell-free network}
    \label{NSP_relay}
  \end{subfigure}
  \vspace{0.1cm} 
  \caption{Two different HAPS-based cell-free networks in \cite{2023_Lou_adhoc}.}
  \label{2023_cellfree}
\end{figure}

\begin{figure}[h]
  \centering
  \begin{subfigure}[b]{1\columnwidth}
    \includegraphics[width=\columnwidth]{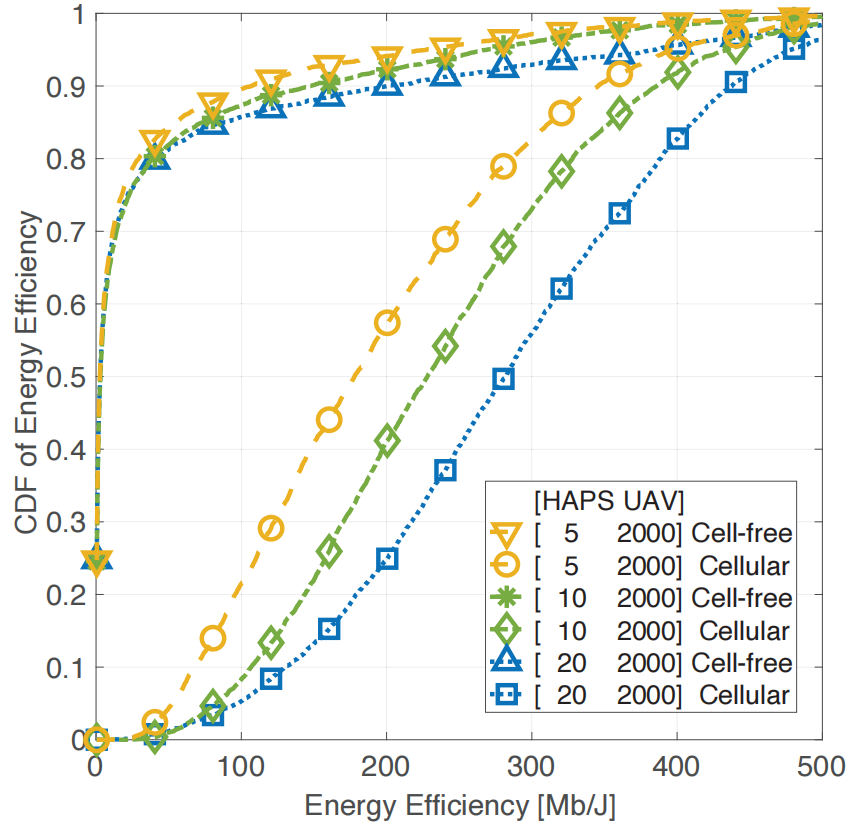}
    \caption{Effect on energy efficiency under different number of HAPS}
    \label{sim_center}
  \end{subfigure}
  \vspace{0.1cm} 
  \begin{subfigure}[b]{1\columnwidth}
    \includegraphics[width=\columnwidth]{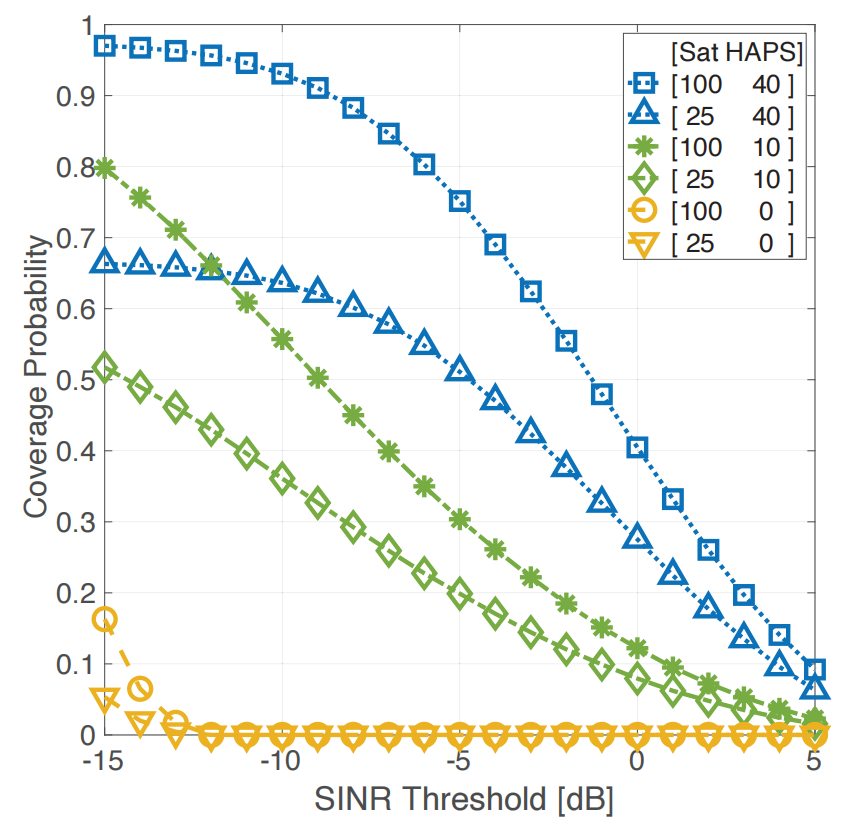}
    \caption{Coverage probability under different number of HAPS}
    \label{sim_relay}
  \end{subfigure}
  \vspace{0.1cm} 
  \caption{The simulation results presented in~\cite{2023_Lou_adhoc}.}
  \label{sim}
\end{figure}

Deploying NSPs for heterogeneous cross-domain networking enables rapid and flexible coverage of targeted areas requiring communication services. D. Xu \textit{et al.} in~\cite{2017_Xu_geomotry} achieved the maximization of coverage rate in different environments and solved the multi-NSP collaborative-deployment problem through iteratively optimising  the NSP spacing and coverage radius. The hybrid FSO/RF multi-hop relaying system proposed in~\cite{2024_Turk_LEO} employed gradient descent to optimize inter-NSP spacing, enabling high throughput and a low outage probability communication under adverse weather conditions. Hierarchical architectures have also gained significant attention. Y. Shibata \textit{et al.} in~\cite{2020_Shibata_GA} designed a hierarchical cellular architecture for NS-ComNet, leveraging hexagonal grid deployment and beam parameter optimization to enhance wide-area spectral efficiency. Hybrid topologies provide universal architectural support for dynamic NS-ComNet through heterogeneous node collaboration.  Two cell-free architectures based on NSPs were proposed in \cite{2023_Lou_adhoc}. In Fig. \ref{2023_cellfree}\subref{NSP_center}, UAVs share information within a HAPS-centered ad-hoc network, reducing resource overhead in distributed routing. In Fig. \ref{2023_cellfree}\subref{NSP_relay}, users establish connections via HAPS functioning as access points in a cell-free architecture, which relays aggregated data to satellites. Fig. \ref{sim}\subref{sim_center} presents the cumulative distribution function (CDF) of different energy efficiencies for different networks, demonstrating the superior energy efficiency of the cell-free network compared to cellular counterparts. Meanwhile Fig. \ref{sim}\subref{sim_relay} reveals that an increase in HAPS deployment density leads to enhanced coverage probability under fixed satellite constellation sizes. A. MS \textit{et al.} in~\cite{2024_Alam_adhoc} proposed a dual-connectivity NS-ComNet architecture that dynamically optimizes airship deployment positions via multi-objective evolutionary algorithms, achieving high coverage while ensuring network robustness and user QoS requirements. The collaborative networking between NSPs and LEO satellites effectively reduces space-ground propagation distances and overcomes coverage-capacity bottlenecks in dense urban and remote areas. K. Mashiko \textit{et al.} in~\cite{2025_Mashiko_LEO} investigated a hybrid FSO/RF SAGSIN system, proposing a joint optimization method for NSP deployment and satellite/NSP coverage regions to serve high-traffic areas dynamically. Addressing limitations of single-NSP deployments, Z. Niu \textit{et al.} proposed a dynamic zenith angle-optimized networking strategy that achieved attenuation reduction, outage probability minimization, and handover frequency suppression through spatiotemporal load balancing, thereby enhancing end-to-end transmission reliability~\cite{2025_Niu_LEO}. These studies demonstrated significant improvements in NSP coverage scalability and operational flexibility across diverse environments through advanced topological design.

\subsection{Resource Optimization for NS-ComNet Efficiency}
To address the challenges of spectrum sharing and interference in the coexistence scenario of multi-altitude networks, researchers have proposed resource management schemes such as dynamic access control and channel allocation optimization. These approaches significantly improve heterogeneous network capacity and resource utilization while ensuring service fairness. Y. Liu \textit{et al.} in~\cite{2009_Liu_adhoc} introduced a two-dimensional state transition-based coverage model, wherein NSP coverage overlaps and access control policies are dynamically adjusted to prioritize blocking probability and maintain service-level fairness in shared spectrum scenarios. M. Guan \textit{et al.}  proposed a standard for minimum mobile-user access distance~\cite{2016_Guan_adhoc}. By combining the dynamic prediction of the number of users and call volume, a hotspot channel allocation mechanism for NSPs was designed. This mechanism can not only improve the channel utilization rate but also achieve reasonable allocation and reduce the call blocking rate.

Cross-domain resource coordination and optimization are a key focus in the heterogeneous networking of NS-ComNet. H. Ahmadinejad \textit{et al.} in~\cite{2022_Ahmadinejad_UAV} presented a dual-layer NSP-UAV network framework with distributed optimization. It used low-complexity subcarrier and power allocation strategies to support flexible deployment, dynamic interference control, and QoS assurance in rural and post-disaster scenarios. A. H. Arani \textit{et al.} in~\cite{2023_Arani_HAPS_UAV} proposed a hybrid method combining deep Q-networks and fixed-point iterations for joint 3D UAV trajectory and channel assignment, achieving both fairness and load balancing under user mobility conditions. Given LEO satellites' limited power and environmental constraints, NSPs assist in relaying data to ground terminals. A. Alsharoa \textit{et al.} in~\cite{2020_Alsharoa_LEO} proposed a three-tier architecture integrating satellites, NSPs, and ground base stations, employing a two-stage joint optimization of NSP placement and resource allocation to improve global user throughput. To efficiently manage the limited power resources in non-terrestrial networks, Z. Ali \textit{et al.} proposed a soft actor-critic RL framework that jointly optimizes power allocation across FSO fronthaul and RF broadcast links~\cite{2024_Ali_HAPS_LEO}. This approach achieved a balance between energy efficiency and throughput under dynamic environmental conditions. However, it assumed static ground user positions, which limited its applicability in realistic scenarios involving highly dynamic user mobility. In contrast, W. Wen \textit{et al.} shifted the research focus from physical-layer communication optimization toward computation-layer resource management~\cite{2025_Wu_SAGSIN}. Targeting dynamic multi-user environments, the authors proposed a multi-NSP collaborative computation offloading method based on graph-theoretic iterative optimization and resource allocation. By decomposing a mixed-integer nonlinear programming problem, the framework jointly optimized task scheduling and compute resource distribution, resulting in significant reductions in overall system latency for SAGSIN. These studies combined  tightly coupled mathematical optimization with dynamic control to overcome inherent resource constraints in non-terrestrial networks.
\subsection{Handover Management for NS-ComNet Seamless Mobility}
\begin{figure}[h]
    \centering
    \includegraphics[width=0.45\textwidth]{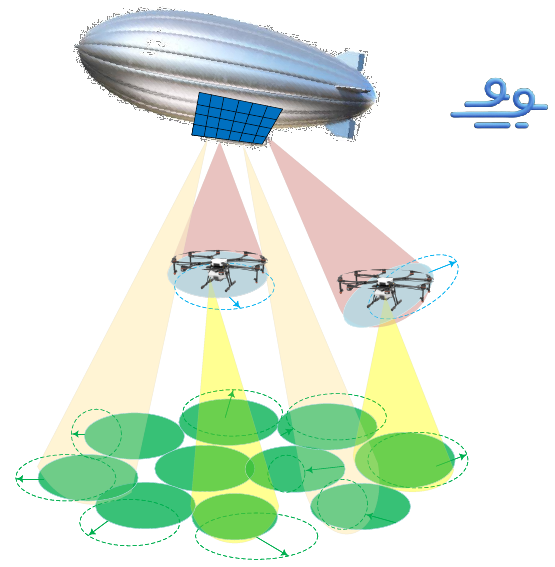}
    \caption{Changes in coverage location or shape caused by airflow and mechanical vibration of NSP.}
    \label{handover}
\end{figure}

In NS-ComNet, the movement of NSPs and atmospheric disturbances can cause changes in their position and orientation. These variations result in dynamic fluctuations in the coverage area and shape, as illustrated in  Fig.~\ref{handover}. Such coverage fluctuations, combined with multi-dimensional mobility of user terminals across domains, data rates, and coverage layers, pose significant challenges for handover management. Traditional handover algorithms developed for terrestrial cellular networks fall short in addressing these complexities, as they are not designed to cope with the dynamic and heterogeneous nature of integrated aerial, space, and terrestrial environments. The absence of coordinated multi-domain handover mechanisms therefore calls for intelligent solutions to ensure link reliability and enhance overall air-space-ground resource efficiency. User-centric handover aims to ensure seamless service continuity during the switching process for users. P. He \textit{et al.} in~\cite{2016_He_handover} proposed an adaptive handover algorithm based on the prediction of signal strength using the improved least mean squares method. By combining the dynamically adjusted detection period with a confidence-interval threshold, the algorithm effectively reduces the number of unnecessary handovers. S.-y. Ni \textit{et al.} in~\cite{2016_Ni_handover} designed a prediction-based handover decision algorithm with adaptive thresholds by considering different required thresholds for received signal strength variation rates, thereby mitigating unnecessary handovers caused by NSP motion-induced fluctuations. The unstable motion of airborne platforms also impacts handover probability. P. He \textit{et al.} in~\cite{2016_He_handover} analyzed the effects of oscillatory states on handovers and derived expressions for average and maximum handover probabilities. S. Alsamhi \textit{et al.} in~\cite{2015_Alsamhi_handover} incorporated the dynamic mobility of NSPs into handover decisions through multi-dimensional data training using radial basis function neural networks. The training considered parameters such as received signal strength, traffic intensity, steerable antenna configuration, elevation angle, delay, bandwidth, HAPS position, and distance to the next base station. This approach significantly reduced handover frequency and call blocking rates while achieving coverage compensation and seamless connectivity.
\subsection{Multi-Objective Joint Optimization for NS-ComNet Performance Trade-off}
Most existing studies~\cite{2024_Alam_adhoc,2012_Zong_adhoc,mashiko2024optimization,2022_Deka_optimization} focus on isolated optimization of communication resources or node deployment, which is insufficient to address the complex requirements of massive terminal access, 3D coverage blind areas, and the guarantee of QoS for multiple services in space networks. There is an urgent need to develop multi-objective joint optimization algorithms to maximize performance while minimizing cost. T. P. Truong \textit{et al.} in~\cite{2022_Truong_MEC} proposed a UAV-based MEC network enhanced by NSPs. A deep deterministic policy gradient-based algorithm was adopted to dynamically optimize both task offloading decisions and offloading rates, enabling intelligent coordination between local computing and NSP resources. In remote areas lacking terrestrial base stations, overall system latency and energy consumption are minimized through the joint optimization of resource scheduling and computation offloading. Y. He \textit{et al.} in~\cite{2022_He_UAV} proposed an alternating optimization framework to jointly design UAV altitudes and spectrum allocation for NSPs-UAV and UAV-user links, employing Hungarian algorithms and successive convex approximation techniques to maximize total uplink/downlink rates. G. Zhang \textit{et al.} in~\cite{2025_Zhang_SAGSIN} introduced a hierarchical cooperative SAGSIN architecture involving a satellite, NSPs, UAVs, and terrestrial IoT devices. Using deep RL, the framework jointly optimized NSP and UAV deployment altitudes as well as coverage areas to achieve end-to-end optimization of wide-area information collection, processing, and transmission under system constraints. Such cross-layer and cross-domain multi-objective joint optimization is crucial to meet future NS-ComNet and SAGSIN requirements for scalability, responsiveness, and resource efficiency.
\subsection{Machine Learning/AI Enabling NS-ComNet Optimization Enhancement}
Traditional mathematical models exhibit limited adaptability and are highly susceptible to environmental variations. In contrast, AI-driven solutions enable autonomous learning of optimal strategies through continuous interaction with the environment, effectively addressing model failures caused by dynamic scenarios. O. Anicho \textit{et al.} in~\cite{2019_Anicho_AI} compared the effectiveness of RL and swarm intelligence in coordinating communication coverage areas of multiple solar-powered UAVs. Swarm intelligence, driven by rule-based logic, exhibits faster convergence and more stable coverage ratios, while RL attains higher peak user coverage, albeit with lower single-NSP coverage due to its exploration strategy. In a similar vein, M. Guan \textit{et al.} in~\cite{2019_Guan_AI} designed a Q-learning-based dynamic channel allocation algorithm that jointly optimizes beamforming and edge-of-cell user parameters through environment-adaptive learning.
This approach enhances coverage efficiency and improves network load balancing in multi-user scenarios. W. Takabatake \textit{et al.} in~\cite{2023_Takabatake_adhoc} proposed a neural network-driven dynamic area optimization algorithm for real-time multi-cell configuration of NSPs under user distribution changes. 

Extending these concepts to more complex scenarios, H. N. Abishu \textit{et al.} developed a blockchain-federated multi-agent deep-RL framework for NSP-assisted vehicular digital twin resource allocation. This framework achieved privacy-preserving system utility maximization while reduced latency and energy consumption~\cite{2024_Abishu_single}. Satellite-terrestrial connectivity is inherently intermittent and irregular, rendering the direct implementation of federated learning (FL) in satellite networks impractical, as training may span several days. To address this, M. Elmahallawy \textit{et al.} introduced NSPs as distributed FL parameter servers within the FedHAP framework, significantly accelerating convergence~\cite{2022_FedHAP_Elmahallawy}. Building upon this, M. Elmahallawy proposed the NomaFedHAP model, which leverages non-orthogonal multiple access to enable concurrent multi-orbit satellite transmissions. It also employs orthogonal frequency-division multiplexing-based techniques to suppress co-orbital interference~\cite{2024_Elmahallawy_LEO_HAPS}. Both theoretical analysis and simulation results demonstrate superior Doppler-shift mitigation, high spectral efficiency, and low outage probability. In the domain of multi-user access control, user mobility in non-terrestrial networks presents unique challenges for handover efficiency and throughput stability. Y. Cao \textit{et al.} in~\cite{2021_Cao_LEO} addressed this by proposing a long short-term memory deep Q-network architecture that captured base station mobility patterns to support autonomous user association decisions. This approach maximized long-term throughput while reducing handover frequency in dynamic topologies. Collectively, these intelligent methods represent a paradigm shift from static configurations toward autonomous, self-optimizing NS-ComNet operations.
	
\section{Future Directions of NS-ComNet in the 6G and Beyond Era}
\label{Future Directions in the 6G and Beyond Era}
As NS-ComNet continues to mature, its integration with emerging paradigms of 6G and beyond becomes both inevitable and indispensable. The unprecedented demands for extremely high reliability, low-latency, and intelligent connectivity in future wireless systems necessitate a reimagining of NS-ComNet architectures and functions. This section outlines the potential directions and research frontiers that will shape the evolution of NS-ComNet in the next-generation communication era. By exploring synergistic technologies such as SDN/NFV orchestration, edge intelligence, advanced antenna systems, and the convergence of THz and optical communications, among others, we aim to highlight key innovations that will drive NS-ComNet towards enhanced autonomy, scalability, and cross-domain collaboration.
\subsection{Global Orchestration of SDN/NFV}
Traditional networks rely on vertically integrated control and data planes. Their complexity and long deployment cycles cannot meet 6G-era demands for agility and personalization. This gap urgently calls for service-oriented, flexible resource management, and on-demand customization. The deep integration of SDN and NFV, achieved through virtualizing network functions (VNF) into building blocks to enhance network reconfigurability, has emerged as the core technical pathway to overcome dynamic management and resource rigidity challenges in VHetNet. SDN enables global resource orchestration through the decoupling of the control and data plane~\cite{2018_Kaur_SDN}. This allows real-time perception of cross-domain topological variations among satellites, NSPs, UAVs, and terrestrial base stations, thereby facilitating dynamic link configuration optimization. As a complement to SDN, NFV implements hardware-software decoupling through VNFs~\cite{2015_Han_NFV}. This transformation enables dedicated hardware to be replaced by dynamically orchestrated software instances, supporting elastic resource scaling, and flexible function deployment.

Network slicing technology further enhances the resource isolation capabilities of SDN and NFV~\cite{2020_Khan_slice}. It abstracts physical network resources into shared virtual resource pools to accommodate services with multi-dimensional requirements. In scenario-specific slicing customization, space-air-ground-sea-integrated slices achieve precise service adaptation through hierarchical QoS assurance and dynamic resource coordination: satellite slices prioritize high-reliability laser backhaul; NSP slices leverage mMIMO beamforming and dynamic spectrum sharing for wide-area coverage; UAV slices employ mmWave links and edge nodes to support low-latency hotspot responses. Building upon this, dynamic slice migration mechanisms reinforce network resilience. In scenarios where wind disturbances create coverage gaps in UAV networks, SDN controllers trigger seamless migration of affected user slices to adjacent NSP nodes based on real-time topology awareness, while NFV ensures service continuity through VNF state snapshot synchronization~\cite{2018_SDNNFV}. This architecture synergizes hierarchical and dynamic mechanisms to satisfy multi-dimensional performance demands across SAGSIN. It also can adaptively counters environmental perturbations to enhance network resilience and improve user experience.

\subsection{MEC, Over the Air Computing, and FL}
Highly real-time services impose stringent requirements for both low latency and high computational power. However, terminal devices often suffer from limited battery capacity and insufficient processing capabilities, creating an inherent contradiction that drives urgent technological innovation. Leveraging wide-area coverage and persistent station-keeping advantages, NSPs enable real-time network state data acquisition and establish near-space computing hubs through onboard MEC, effectively alleviating terrestrial computational burdens. Deploying edge computing nodes in NS-ComNet breaks the limitations of traditional space-ground separation computing, providing millisecond-level low-latency services for satellites, UAVs, and terrestrial devices. Particularly during terrestrial network outages or congestion, this architecture offloads computation-intensive and latency-sensitive tasks to NSP nodes. This approach simultaneously mitigates communication link congestion and enables real-time data processing. However, stand-alone terrestrial-, aerospace-, or space-based MEC systems remain constrained by node mobility, resource limitations, and connection instability, necessitating exploration of multi-node coordination mechanisms. Existing studies have validated enhanced multi-antenna communication offloading efficiency in LEO satellite-NSP collaborative architectures~\cite{2022_Ding_MEC}, yet its dynamic-coordination adaptability in complex offloading scenarios requires further optimization. 

Confronting dynamic network environment challenges, traditional cloud computing exhibits excessive latency and energy consumption due to long-distance transmissions. Integrating edge intelligence enables NSP to perform local inference, task scheduling, and adaptive resource control with minimal delay. Recent advances have further proposed communication-efficient federated edge learning (FEEL) frameworks that embed intelligence directly into the air interface. Specifically, by leveraging the signal superposition property of the wireless medium, model parameters can be aggregated or partially computed during the transmission process itself, eliminating the need for conventional decoding and post-processing. Notably, Qiao \textit{et al.} introduced a massive digital over-the-air computation scheme that combined vector quantization with grant-free massive access and approximate message passing, achieving efficient model aggregation across a large number of AI of things devices~\cite{2024_Qiao_CNSC}. Their approach not only accelerated FEEL convergence with minimal communication overhead but also maintained robustness under low signal-to-noise ratio and asynchronous transmission conditions, offering strong theoretical convergence guarantees and practical scalability under realistic wireless constraints.

\subsection{Breakthroughs in Massive MIMO and Advanced Antenna Array Technologies}
In non-terrestrial networks (NTN), communication links often span tens to thousands of kilometers, where the space-air-ground channels suffer from severe free-space path loss and signal attenuation, posing fundamental constraints on system performance. The mMIMO technology, with its highly directional beamforming capability~\cite{2023_Gao_ISAC}, offers physical-layer support to compensate for the excessive path loss inherent in long-range space-terrestrial links. By deploying high-dimensional antenna arrays on NSPs, transmit energy can be spatially focused to achieve long-distance, high-precision coverage, thereby significantly improving the link budget and enhancing communication reliability under stringent power constraints.

Meanwhile, the advanced antenna array structure with high beam guidance and spatial resolution provides a physical basis for dynamic user tracking, interference suppression, and intelligent beam scheduling, thereby countering high path loss. For example, the fluid antenna system \cite{2024_FAS_WKK}, which has a dynamically reconfigurable conductive structure, achieves quasi-continuous angular scanning and adaptive beam alignment by adjusting the shape and position of the antenna. RIS adopts many passive and low-cost reflective components, has programmable phase shift, and can intelligently reconfigure the signal propagation path \cite{2024_ykk_RIS,2023_ran_RIS}. Furthermore, the stacked intelligent meta-surface integrates a multi-layer programmable meta-structure \cite{2024_SIM} in the 3D configuration, achieving hierarchical beam construction and energy focusing, and enhancing spatial control. Despite the significant progress achieved in terrestrial networks, deploying mMIMO on NSPs remains highly challenging. From a hardware perspective, the severe payload constraints, stratospheric environmental conditions, and aerodynamic limitations impose strict requirements on the lightweight and weather-resilient design of antenna arrays.

\begin{figure}[h]
    \centering
    \includegraphics[width=0.45\textwidth]{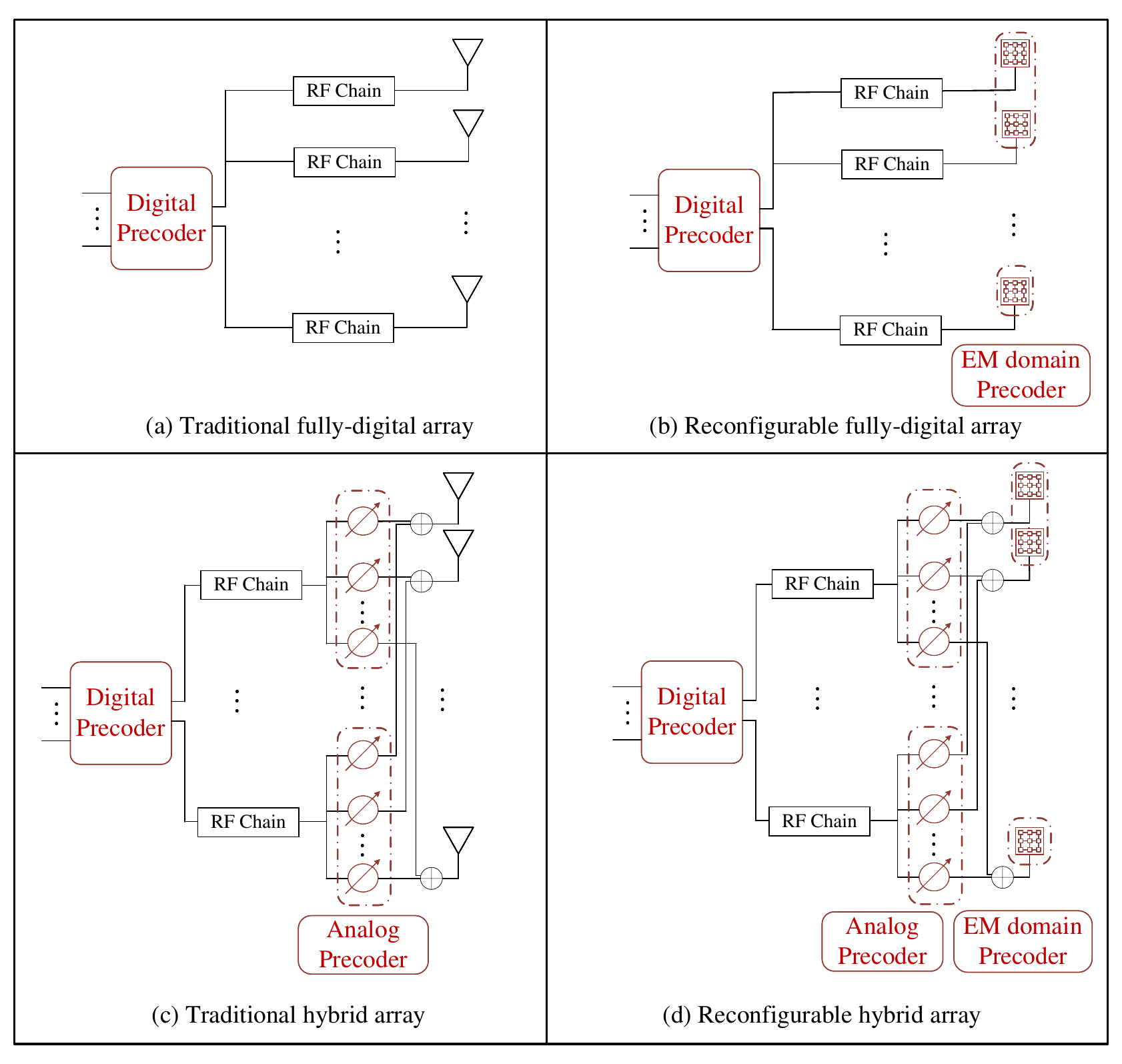}
    \caption{Comparison between traditional fully-digital/hybrid arrays and their reconfigurable counterparts.}
    \label{MIMO}
\end{figure}

To alleviate these limitations, K. Ying \textit{et al.} in~\cite{2024_Ykk} proposed a reconfigurable massive MIMO (RmMIMO) architecture that introduced a three-stage hybrid precoding framework in the electromagnetic (EM) domain. As illustrated in Fig.~\ref{MIMO}(d), this architecture was constructed by replacing the antennas in a conventional hybrid array (Fig.~\ref{MIMO}(c)) with reconfigurable antennas. The resulting structure integrated digital-domain precoding at the baseband, analog-domain precoding at the RF stage, and radiation pattern precoding at the EM level. This architecture enabled dynamic adjustment of the radiation pattern for each antenna element. It extended the degrees of freedom from the spatial domain to the electromagnetic domain. Even under a limited aperture budget, the system supported efficient beamforming optimization through manifold-based methods. The approach enhanced spatial flexibility. It also improved energy efficiency by reducing reliance on bulky hardware. Furthermore, it offered better adaptability to dynamic near-space scenarios with fast-varying channels. Compared with traditional fully digital arrays (Fig.~\ref{MIMO}(a)) and reconfigurable fully digital arrays (Fig.~\ref{MIMO}(b)), the proposed EM-domain hybrid RmMIMO architecture demonstrated significantly improved spectral efficiency. It provided a practical and scalable solution for realizing large-scale antenna arrays on size- and weight-constrained NSP.
\subsection{THz-Optical Communication Convergence}
To unleash the performance potential of NS-ComNet, the deep integration of THz and FSO communications is emerging as a key pathway to overcome capacity and coverage bottlenecks. THz technology leverages ultra-wide bandwidth and low atmospheric attenuation in the near-space environments to support hundreds of Gbps data-rate transmissions for beyond LoS NSP interconnections~\cite{2021_THz}. However, it faces deployment challenges such as atmospheric refraction effects, beam misalignment caused by platform micromotion and nonlinear device distortion. FSO demonstrates Tbps-level transmission potential in low-turbulence near-space environments, yet remains constrained by sky radiation noise and transmitter power limitations~\cite{2014_FSO}. The synergistic combination of THz and FSO enables a complementary architecture in NS-ComNet, where THz links can serve as medium-to-long-range high-capacity backbones with moderate alignment requirements~\cite{2023_THz/FSO}, while FSO provides ultra-high-throughput enhancement in short-range or LoS scenarios. The hybrid system optimizes beamforming via large-scale antenna arrays and suppresses noise through adaptive optical modulation~\cite{2008_CDMA}. In deep-space or inter-satellite applications, FSO remains the dominant solution for long-distance communication due to its narrow-beam efficiency and minimal attenuation in vacuum. Future progress in FL and digital twin-enabled cross-domain resource orchestration will further drive the seamless fusion of THz, optical, and RF multi-band, ultimately establishing a space-air communication paradigm characterized by ultra-high spectral efficiency and all-weather-reliable connectivity.

\subsection{Quantum-Classical Hybrid Computing Security Enhancement}
The convergence of quantum and classical technologies provides a multi-layered security framework to address complex safety challenges in SAGSIN. These challenges include signal leakage and malicious interception in high-frequency wireless transmissions due to narrow-beam misalignment or atmospheric turbulence, as well as data tampering risks in long-range aerial links within NS-ComNet. By synergizing post-quantum cryptography with quantum-secured key distribution, this hybrid approach establishes cross-domain security architectures~\cite{Chu_2021}. For example, NSPs equipped with quantum relaying capabilities enable ultra-long-distance key negotiation between satellites and terrestrial stations~\cite{2021_QuantumSatellite}. Meanwhile, the integration of distributed machine learning with quantum communication technologies helps prevent model data leakage under privacy-preserving constraints. To mitigate the risks of node compromise in NSPs, quantum true random number generation~\cite{2016_Quantum}, coupled with the immutability of blockchain, ensure both data transmission integrity and storage reliability. Furthermore, deep integration of quantum technologies and AI enables real-time anomaly detection, optimizing communication protocols against threats such as mmWave signal leakage. Looking ahead, quantum-classical hybrid security systems are expected to overcome environmental noise and hardware limitations through quantum error correction and novel cryptographic protocols. These systems aim to establish comprehensive protection encompassing encryption algorithm resilience, global key management, and protocol trust verification, thereby laying a secure foundation for SAGSIN operations.
\begin{figure*}[h]
    \centering
    \includegraphics[width=1\textwidth]{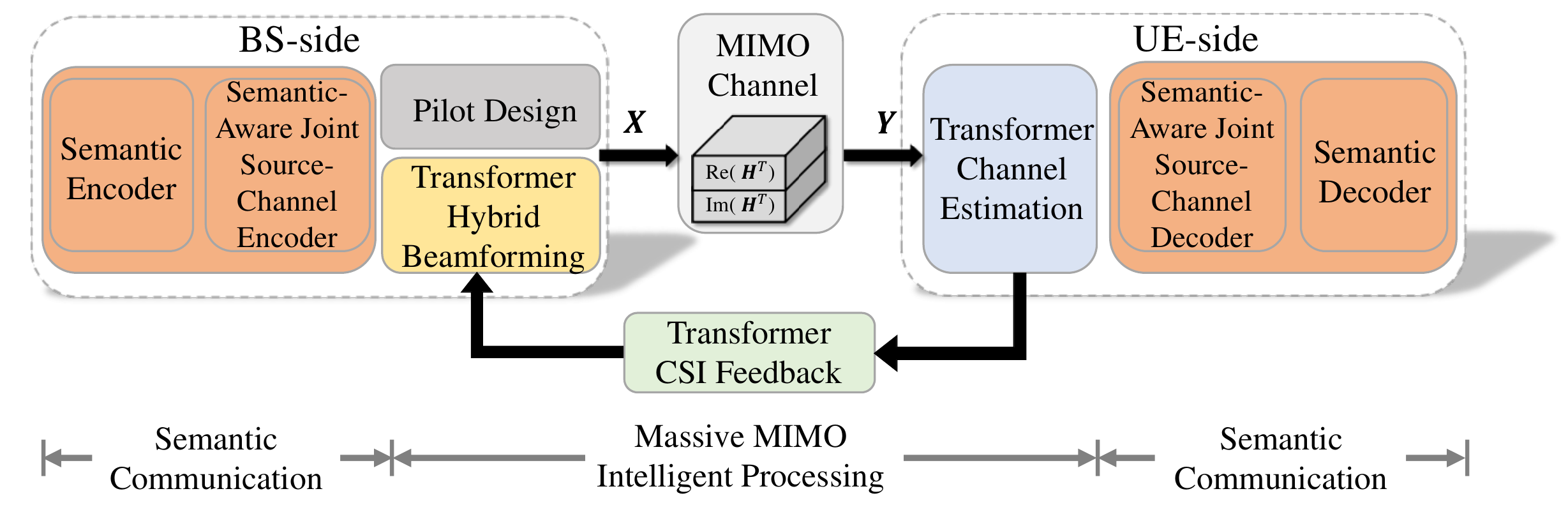}
    \caption{Transformer-based signal processing architecture for mMIMO systems in~\cite{2023_yang_AI}.}
    \label{LargeModel}
\end{figure*}
\subsection{Direct NSP-to-Device Communications}
As a novel access paradigm within NTN, direct NSP-to-device communications (DNDC) refers to user equipment establishing data links with high-altitude platforms without relying on terrestrial relay infrastructure. Compared to satellite-based systems\cite{2025_wang_access}, NSP offers advantages such as lower communication latency, greater deployment flexibility, and cost-efficiency, making it particularly suitable for rapid network provisioning in scenarios such as emergency response, remote area connectivity, and medium-to-low capacity broadband access. Depending on the communication capabilities of the platform and system architecture, and referring to the space-terrestrial integrated network architecture~\cite{2024_He_DSDC}, DNDC can also be divided into three representative technical routes: (i) Dual-mode terminal access, where dedicated NSP communication modules or protocol stacks are embedded into smartphones, enabling direct connectivity via customized waveforms and air interfaces, (ii) Unmodified terminal access, where the NSP is equipped with regenerative payloads capable of handling air interface signaling and protocol processing, thus allowing legacy smartphones to connect without hardware modifications, (iii) Standardized integrated access, which incorporates NSP nodes into the unified air interface and core network architecture defined by 3rd Generation Partnership Project NTN standards, enabling seamless interoperability with terrestrial cellular and satellite systems. Future research efforts should focus on overcoming key technical challenges, including high-precision NSP positioning and attitude control, spectrum coexistence and interference mitigation, link budget modeling and adaptive optimization for low-gain terminals, as well as cross-domain resource orchestration and multi-hop collaborative scheduling in SAGSIN. These advancements will significantly extend the spatial dimension and application scope of NTN-enabled services in the 6G era.

\subsection{AI-Enabling NS-ComNet Physcial-Layer Transmission: From Task-Specified Model to Large Model}
Current task-specified models are typically designed as combinations of specialized processing blocks, such as channel estimation, equalization, and coding or decoding. However, in the physical-layer design of NS-ComNet, traditional model-based modular approaches face increasing challenges. These include high-dimensional processing, complex environments, and rapidly changing interference. Such methods can hardly meet the extreme requirements of future networks in terms of ultra-high capacity, ultra-low latency, and ubiquitous intelligence. In recent years, deep learning has emerged as a promising candidate for 6G physical-layer design. Its powerful feature extraction and nonlinear modeling capabilities provide new possibilities for complex signal processing. Among various architectures, the transformer has shown excellent performance in multi-task communication systems. This is due to its self-attention mechanism and parallel multi-head representation, which support cross-domain modeling and high scalability. At the current stage, researchers usually train lightweight models for specific tasks, such as channel estimation, channel state information (CSI) feedback~\cite{2024_Kui_CE}, and hybrid beamforming~\cite{2022_Wu_CSI}. However, these task-specific models lack generalization and cannot support joint optimization in multi-modal and multi-scenario settings.

To address this issue, Y. Wang \textit{et al.} proposed a unified transformer-based processing framework for 6G intelligent communication in~\cite{2023_yang_AI}. This framework enabled end-to-end optimization from sensing and channel estimation to CSI compression and beamforming. In addition, the transformer was used in the semantic communication module for semantic abstraction and reconstruction. As shown in Fig.~\ref{LargeModel}, the system deployed transformer modules at both the base station (BS) and user equipment (UE). This enabled a unified model loop between the mMIMO processing chain and the semantic communication chain. Furthermore, in the transition toward large model, the transformer was embedded as a core backbone in NSP intelligent agents. It integrated environmental perception, link state, and semantic objectives. This allowed the system to evolve from bit-level transmission to semantic-level and intent-aware communication. Such a paradigm simplified physical-layer signal processing, improved model generalization, and builded a unified foundation for future capabilities such as multi-modal communication, cross-band coordination, and task-adaptive resource scheduling.

Looking ahead, large model is expected to play a central role in NS-ComNet. They will support adaptive learning across multiple tasks and drive the shift from task-specific models to shared multi-task models. This will enable intelligent communication across dynamic and heterogeneous NS-ComNet.

\subsection{AI-Driven Semantic Communication Paradigm Innovation}
Semantic communication emerges as a groundbreaking technology for NS-ComNet. It addresses the contradiction between constrained spectral resources and massive data transmission demands through deep feature extraction and fusion encoding of information semantics~\cite{2024_Wu_Semantic}. This approach enables intelligent, channel-adaptive transmission with substantial data compression. It can further integrates mMIMO beamforming to achieve two key enhancements: semantic-feature-augmented cross-channel interference suppression, which overcomes traditional ``cliff effect" limitations; and end-to-end joint optimization mechanisms that tightly couple semantic fidelity with physical-layer transmission efficiency. Under complex near-space channel conditions characterized by high-speed mobility and time-varying multi-path effects, this architecture supports high-quality multi-user image reconstruction~\cite{2024_Qiao_Semantic}. It thereby establishes a SAGSIN-compatible intelligent communication paradigm that simultaneously delivers ultra-high efficiency, transmission robustness, and mission-aware adaptability.
\subsection{Integration of Communication-Navigation-Sensing-Computing in SAGSIN and Low-Altitude Economy}
\begin{figure}[h]
    \centering
    \includegraphics[width=0.5\textwidth]{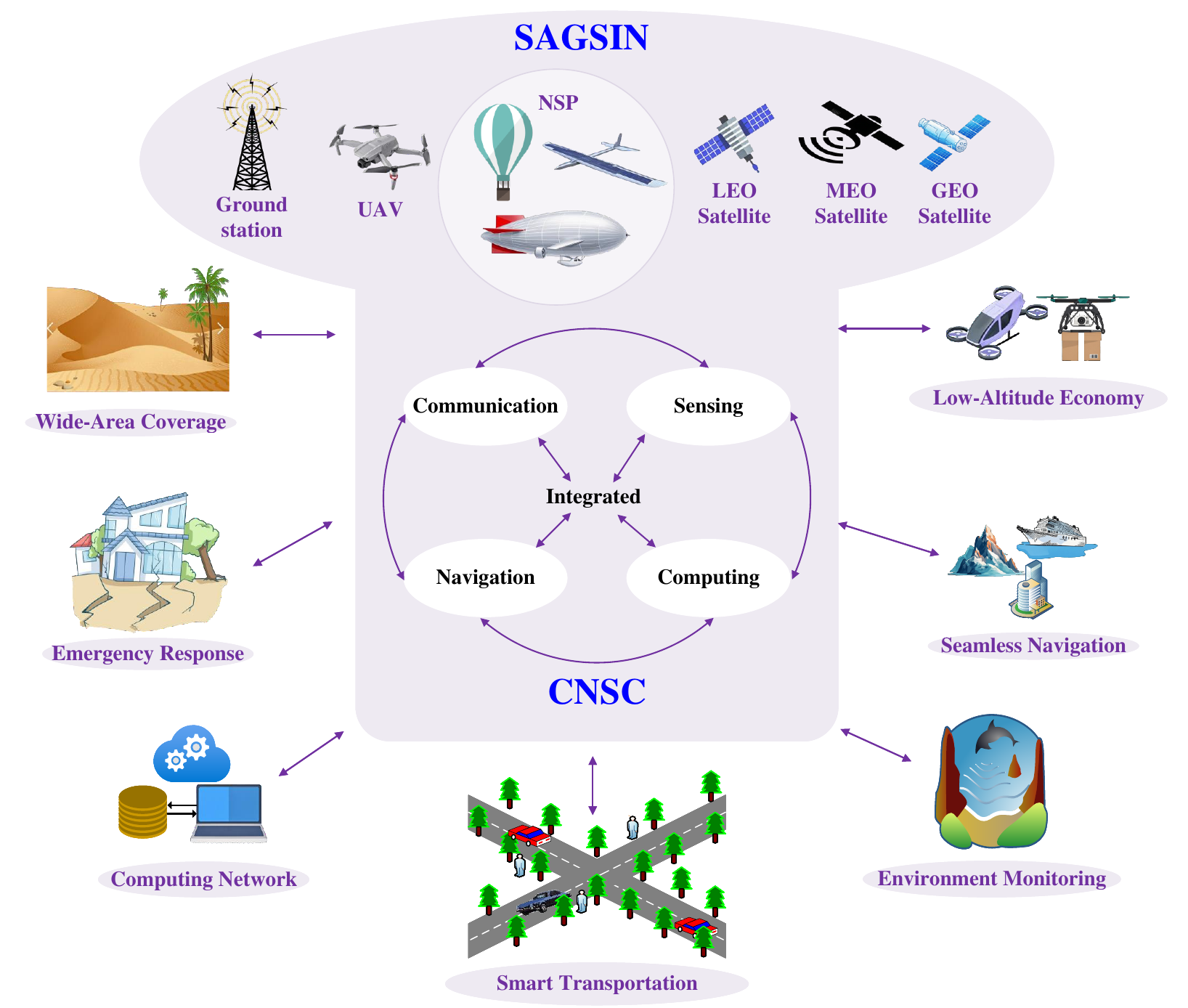}
    \caption{The integration of CNSC technology.}
    \label{CNSC}
\end{figure}

The integration of communication-navigation-sensing-computing (CNSC) technology, as shown in Fig.~\ref{CNSC}, endows NS-ComNet with multi-dimensional synergistic capabilities. Navigation-assisted dynamic topology optimization of communication nodes, combined with sensing-driven real-time channel parameter calibration, significantly reduce transmission latency and path loss~\cite{2025_Gao_ISAC, 2024_Qiao_CNSC}. Simultaneously, the deep integration of MEC with communication protocols enables both intelligent spectrum sharing and cryptographic acceleration in complex scenarios, effectively overcoming resource constraints at NSPs. This converged architecture empowers the rapid deployment of space-terrestrial intelligent relay networks for emergency response and wide-area monitoring. It leverages multi-source sensing data to dynamically coordinate UAV-NSP collaborative links, implementing predictive cross-domain resource scheduling via AI~\cite{2024_fei_ISAC}. Its core value lies in transcending traditional layered design paradigms by unifying physical-layer transmission efficiency, network-layer resilience, and mission-layer intent comprehension into a self-consistent space-air environment framework. This unification establishes the technical foundation for intelligence-driven networking in near-space domains in the 6G and beyond era.
\section{Conclusions}
NS-ComNet, as a core pillar for achieving seamless global coverage in next-generation wireless communications, is propelling the evolution of SAGSIN. By comparing the differentiated characteristics of near-space with other aerial domains, this paper highlights the unique value of NSPs in dimensions such as wide-area persistent hovering and dynamic topology reconfiguration. Furthermore, it systematically reviews recent research advances and industrial practices in the near-space networking technologies, focusing on core challenges and key enabling technologies, and provides an in-depth analysis of their multi-dimensional collaborative potential as space-air hubs. Additionally, this paper prospectively discusses emerging technologies that can enhance the capabilities of NS-ComNet through software-defined, intelligent, and energy-efficient upgrades. Research demonstrates that NS-ComNet, as a crucial nexus for cross-domain multi-dimensional collaboration within SAGSIN, will underpin the ubiquitous connection requirements of 6G scenarios through its wide-area coverage and resilient networking capabilities.

\bibliographystyle{IEEEtran}
\bibliography{sample}

\end{document}